\documentclass[12pt]{iopart}

\usepackage{graphicx}
\usepackage[T1]{fontenc}
\expandafter\let\csname equation*\endcsname\relax
\expandafter\let\csname endequation*\endcsname\relax
\usepackage{amsmath}       
\usepackage{epsfig} 
\usepackage{ulem}
\usepackage{amssymb}
\usepackage[mathscr]{eucal}
\usepackage[hidelinks,colorlinks]{hyperref}
\usepackage[dvipsnames]{xcolor}
\definecolor{ao(english)}{rgb}{0.0, 0.5, 0.0}
\usepackage{bbold}
\usepackage{placeins}
\usepackage{bm}
\usepackage{empheq}
\usepackage{caption} 
\usepackage{etoolbox}
\makeatletter
\newcommand{\mainmatter}{%
  \setcounter{footnote}{0}%
  \patchcmd{\@makefntext}{\fnsymbol}{\arabic}{}{}%
  \patchcmd{\@thefnmark}{\fnsymbol}{\arabic}{}{}%
  \def\@makefnmark{\textsuperscript{\arabic{footnote}}}%
}
\makeatother

\newcommand{\tauFWHM}{\tau_{\rm\scriptscriptstyle{FWHM}}}

\newcommand{\Smilei}{{\sc Smilei}\,}

\newcommand*\widebar[1]{%  %%% Warning: weird stuff happens if I remove the seemingly-useless '%'
   \vbox{%
     \hrule height 0.5pt%    % Line above with certain width
     \kern0.15ex%          % Distance between line and content
     \hbox{%
       \kern-0.2em%   % Distance between content and left side of box, negative values for lines shorter than content
       \ifmmode#1\else\ensuremath{#1}\fi%  % The content, typeset in dependence of mode
       \kern-0.1em%                        % Distance between content and left side of box, negative values for lines shorter than content
     }% end of hbox
   }% end of vbox
}

\begin{document} 

%%% TITLE %%%%%%%%%%%%%%%%%%%%%%%%%%%%%%%%%%%%%%%%%%%%%%%%%%%%%%%%%%%%%%%%%%%%%%%%%%%

\title{Impact of the laser spatio-temporal shape on %nonlinear
Breit-Wheeler pair production}

\author{A. Mercuri-Baron$^1$, M. Grech$^2$, F. Niel$^1$, A. Grassi$^1$, M.Lobet$^3$, A. Di Piazza$^4$ and C. Riconda$^1$}
\address{$^1$ LULI, Sorbonne Université, CNRS, CEA, École Polytechnique, Institut Polytechnique de Paris, F-75255 Paris, France}
\address{$^2$ LULI, CNRS, CEA, Sorbonne Universit\'{e}, École Polytechnique, Institut Polytechnique de Paris, F-91128 Palaiseau, France}
\address{$^3$ Maison de la Simulation, CEA, CNRS, Universit\'{e} Paris-Sud, UVSQ, Université Paris-Saclay, F-91191 Gif-sur-Yvette,France}
\address{$^4$ Max  Planck  Institute  for  Nuclear  Physics,  Saupfercheckweg  1,  D-69117,  Heidelberg,  Germany}
\ead{anthony.mercuri@sorbonne-universite.fr}

\begin{abstract}

The forthcoming generation of multi-petawatt lasers opens the way to abundant pair production by the nonlinear Breit-Wheeler process, i.e., the decay of a photon into an electron-positron pair inside an intense laser field. In this paper we explore the optimal conditions for Breit-Wheeler pair production in the head-on collision of a laser pulse with gamma photons. The role of the laser peak intensity versus the focal spot size and shape is examined keeping a constant laser energy to match experimental constraints. A simple model for the soft-shower case, where most pairs originate from the decay of the initial gamma photons, is derived. This approach provides us with a semi-analytical model for more complex situations involving either Gaussian or Laguerre-Gauss (LG) laser beams. We then explore the influence of the order of the LG beams on pair creation. Finally we obtain the result that, above a given threshold, a larger spot size (or a higher order in the case of LG laser beams) is more favorable than a higher peak intensity. Our results match very well with three-dimensional particle-in-cell simulations and can be used to guide upcoming experimental campaigns.
\end{abstract}

\maketitle
\mainmatter %%% to enable numbered footnotes

\section{Introduction}

Abundant electron-positron pair production is among the most exotic and striking processes of extremely high-intensity laser interaction with light and/or matter that will be made possible by the emerging class of multi-petawatt (multi-PW) laser systems \cite{Di_Piazza2012}. Under conditions envisioned for forthcoming facilities such as Apollon~\cite{Apollon}, CoReLS~\cite{corels}, ELI~\cite{ELI}, OMEGA-EP OPAL~\cite{Bromage_2019}, XCELS~\cite{XCELS} and {\sc zeus}~\cite{ZEUS},
the dominant process leading to pair production is nonlinear Breit-Wheeler
, i.e. the decay of a high-energy photon, as it interacts with the strong laser field, in an electron-positron pair~\cite{Reiss_1962,Nikishov1964}.

In the last decade, various simulation campaigns have been conducted to help design the future experiments to efficiently drive this process, an effort that was largely made possible and supported by particle-in-cell (PIC) simulations accounting for quantum electrodynamics (QED) processes  \cite{Duclous2010, Elkina2011, Arber2015, Gonoskov2015, Grismayer2016, Lobet2016}. 
Several configurations have been considered. Electron-seeded electromagnetic/QED cascades in the collision of two counter-propagating laser pulses, a setup originally proposed by Bell \& Kirk 
\cite{Kirk2008}, have attracted particular attention \cite{Elkina2011, Grismayer2016,Nerush2011,Grismayer2017,Tamburini_2017,Jirka2017} as such cascades were identified as a possible limitation on the attainable intensity of high power lasers \cite{Fedotov2010}. This setup was further built upon considering either the use of plasma channels~\cite{Zhu2019} or the collision of multiple laser pulses~\cite{Gelfer2015,Vranic2016,Gonoskov2017}. All these works considered multiple laser beams, but electromagnetic cascades were also predicted considering the direct irradiation of a solid target by a single, extremely intense laser pulse, a situation that leads to the production of dense pair plasmas \cite{Ridgers2012} (see also \cite{Lobet2015} for an application to laboratory astrophysics). In these cases, the solid target, an overdense plasma acts as a mirror and pair production is efficiently achieved in the field of the incident and reflected laser light at the target front~\cite{Kostyukov2016}.

All the previously discussed configurations considered cascades seeded by electrons initially at rest in the laboratory frame. The seed electrons, strongly accelerated in the electromagnetic fields of laser pulses with extreme intensity (typically $\sim 10^{24}\, {\rm W/cm^2}$), gain energies at the GeV-level (as expected, at such intensities, from the ponderomotive scaling) and are the source (via nonlinear Compton scattering) of the high-energy photons triggering pair production. 
Another possible path toward pair production in the laboratory relies on an already existing high-energy electron or photon source. The two latter schemes are related, as photons with energy of the order of the electron one are produced via nonlinear Compton scattering of electrons in the intense laser fields~\cite{Mackenroth2011, blackburn2014, Niel2018}.
Indeed, this configuration was exploited in the seminal E-144 experiment at SLAC~\cite{burke}, where the head-on collision of 46.6 GeV electron beams with TW laser pulses gave the first direct demonstration of nonlinear Breit-Wheeler pair production. However, the laser intensity in this experiment was only marginally relativistic, and very few positrons were produced ($\sim 100$ positrons over nearly $22000$ electron-laser collisions). %%%Additional experiments are planned in this configuration at both {\sc luxe}~\cite{abramowicz2019} and {\sc facet-ii}~\cite{Yakimenko2019} facilities, where the TW-class laser intensity and energy are substantially higher than in this seminal work of Burke et al. 

Combining GeV-electron beams with PW and multi-PW lasers will allow to extend this study deep into the relativistic regime, as investigated in~\cite{Lobet2017}
that numerically demonstrated the possibility of producing extremely bright high-energy photon sources as well as positrons bunches for laser intensities $\gtrsim 10^{22} {\rm W/cm^2}$, in conditions relevant to the Apollon facility.
This configuration was further studied theoretically and via PIC simulations in ~\cite{Blackburn2017,Vranic2018,Chen_Keitel2018}, while the possibility to produce pairs via nonlinear Breit-Wheeler with high-energy photons from a Bremsstrahlung source was presented in~\cite{Blackburn2018}.\\

In this paper, we focus on the study of nonlinear Breit-Wheeler pair production following from the head-on collision of an extremely intense laser (intensities in the range $10^{21}-10^{25} {\rm W/cm^2}$) with a burst of gamma photons (with energies ranging from 100 MeV to few 10s of GeV). In contrast with~\cite{Blackburn2018}, where the authors focused on the optimization of the photon source, we consider here the high-energy photon burst as given (only its energy will vary), and do not discuss its origin (various sources of high-energy photons have been proposed \cite{Phuoc2012,Gong2017,Vranic2019,Magnusson2019,Zhu2020,Sampath2021}). Rather, we aim to optimize the conditions of interaction with the colliding high-intensity laser. Motivated by experimental constraints, we investigate the optimal conditions for pair production varying the laser polarisation, focusing, spatial or temporal profiles, always considering a fixed laser energy. 

In particular, special attention will be paid to the use of Laguerre-Gauss (LG) beams, which have recently attracted the interest of the ultra-high intensity laser-plasma community~\cite{Chen_Keitel2018,Vieira2014,Nuter2020,Longman2020,Duff2020} for their ability to carry orbital angular momentum (OAM)~\cite{allen}. These beams, that emerge as eigen-modes of the paraxial equation, have unique properties, such as a ring-shaped intensity distribution and %the fact that they can carry 
the large OAM, that could have an impact on pair production. It was for instance demonstrated in~\cite{Chen_Keitel2018} that the collision of an electron beam with a LG beam can lead to efficient production of gamma rays carrying large OAM, to enhanced secondary radiation emission and pair production. However, that study was performed at constant maximum intensity, so it is hard to distinguish the impact of the increased energy (up to $3\times$ higher for the LG beam) from the role of the laser spatial profile itself. We will discuss this aspect in detail in this paper.\\

Because of the remarkable number of free parameters to study (polarisation, focusing, spatial and temporal profile) and the complex spatio-temporal dependence of the electromagnetic fields of LG beams, it is useful to develop and validate a reduced analytical model for pair creation (see Ref. \cite{Di_Piazza_2016} for an analytical study of the Breit-Wheeler pair production process in a Gussian beam). Starting from the probability for a given high-energy photon to decay in an electron-positron pair while it crosses an extremely intense laser pulse, we propose 
a simple, yet accurate, model to compute the number of pairs 
created during this interaction. The computation relies on three assumptions.
(i) The electron-positron pair production rates are obtained in the locally constant cross-field approximation \cite{Meuren2015}.
(ii) The newly created charged particles do not further contribute to pair production. 
(iii) The potentially complex electromagnetic fields explored by the high-energy photons as they cross the laser pulse are approximated, assuming that each photon sees a succession of single half-periods of monochromatic plane waves, the field amplitude of each half-period being computed so that both the temporal and spatial profiles of the laser pulse at focus are accounted for. As will be further discussed in this work, these assumptions are well satisfied for a wide range of parameters of interest for this study and currently accessible laser parameters. 
Our model goes further than previous studies \cite{Blackburn2017,Meuren2015} as it accounts for arbitrary polarisation and spatio-temporal profile of the laser beam, and it is not limited to small pair production probability and/or rate.  

Three-dimensional PIC simulations performed with the code \Smilei~\cite{smilei} and embarking the relevant QED modules~\cite{Lobet2016} are presented and validate our model. 
Assumptions (ii) and (iii) are shown to hold under conditions relevant to ultra-short laser pulses (up to few tens laser cycles) typical of the Ti:Sapphire technology.
Assumption (ii) holding is of particular importance as it demonstrates that, under conditions relevant to forthcoming ultra-short laser facilities, the high-energy-photon-laser interaction proceeds essentially in a regime which we refer to as the soft-shower regime~\cite{fedotov-shower-cascade}.
Furthermore, our PIC simulations allow to test our model against complex laser pulse geometries, e.g. considering LG beams, showing that assumption (iii) - which is at the core of the simplification of our model - provides accurate predictions even for non trivial electromagnetic field configurations.

The proposed model and performed study make it possible to clearly identify the laser and high-energy photon parameters that optimize pair production for a given laser energy. This work also highlights the conditions in which either stronger focusing or increased gamma photon energy does not improve the production of primary pairs. \\

The paper is structured as follows. In Sec.~\ref{sec:reducedModel}, we introduce the key properties of the nonlinear Breit-Wheeler process, and we propose a reduced model for pair creation in the head-on collision between gamma photons and a plane wave with arbitrary polarization. The possibility to account for a time envelope of the laser field is also considered and the model is benchmarked against 1D particle-in-cell (PIC) simulations. In Sec.~\ref{sec::GeomEffects} we consider the effect of the spatial dependence of the laser field. Both Laguerre-Gauss and Gaussian beams are considered and we introduce the notion of a total cross-section for which we propose a reduced model, shown to be in very good agreement with 3D PIC simulations. The optimal conditions for pair creation with respect to the laser geometry and intensity are also discussed. In Sec.~\ref{sec:discussion} we then use our model to discuss pair production on upcoming Ti:Sapphire laser facilities, before giving our conclusions in Sec.~\ref{sec:conclusions}.

\section{Reduced model for pair production in a time-varying background field}\label{sec:reducedModel}

This section provides the theoretical framework for describing nonlinear Breit-Wheeler pair production, and aims to establish a simple model for pair production in a time-varying background field. Throughout this section, we consider the head-on collision of high-energy gamma photons with a strong, optical, laser pulse (also referred to as the background field). In addition to exact expressions, we provide practical formulae relevant to forthcoming high-power laser facilities, focusing on optical laser systems with micrometric wavelength $\lambda$, with relativistic field strength $a_0 = e E_0/(m_e c\omega) \gg 1$ (with $\omega=2\pi c/\lambda$ the background field angular frequency), corresponding to $I_0 \lambda^2$ well beyond $10^{18} {\rm W/cm^2\,\mu m^2}$, with $I_0 = \epsilon_0 c E_0^2/2$ the laser intensity.
Note that SI units are used throughout this work: $e$, $m_e$ and $c$ denote the elementary charge, electron mass and speed of light in vacuum, respectively, $\epsilon_0$ the vacuum permittivity, $\tau_e=e^2/(4\pi\epsilon_0 m_e c^3)$ the time for light to cross the classical radius of the electron, $\hbar$ the reduced Planck constant and $\alpha = e^2/(4\pi\epsilon_0 \hbar c)$ the fine-structure constant.

\subsection{Pair production instantaneous rate and probability}

\begin{figure}\centering%%%
    \includegraphics[width=0.9\textwidth]{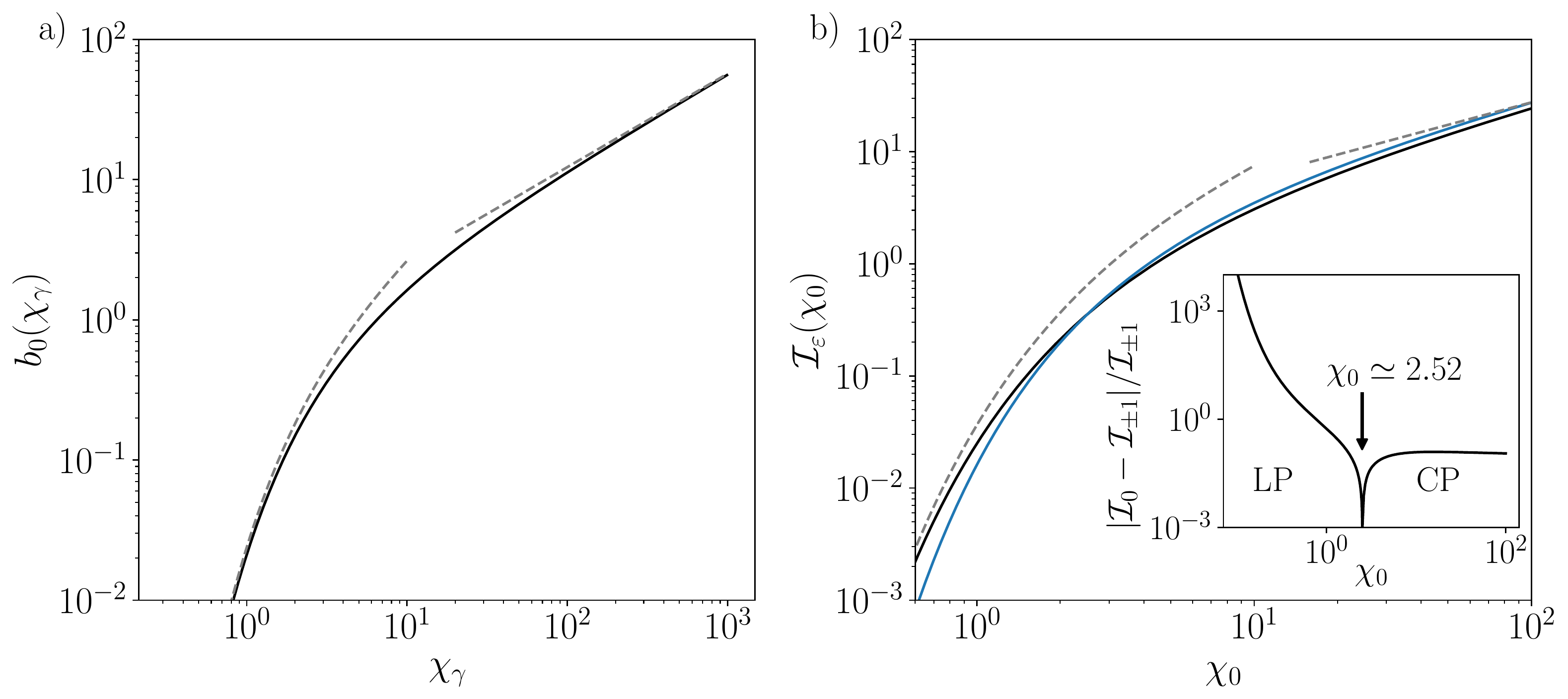}
    \caption{
    (a) Dependence of $b_0(\chi_{\gamma})$ on the photon quantum parameter $\chi_{\gamma}$ as defined by Eq.~\eqref{eq:b0} (black line) and asymptotic behaviors given by Eqs.~\eqref{eq:b0smallChi} and~\eqref{eq:b0largeChi} (grey dashed lines).
    (b) Dependence of $\mathcal{I}_{\varepsilon}(\chi_0)$ [Eq.~\eqref{eq:intI}] 
    on the maximum photon quantum parameter $\chi_0$
    [as defined by Eq.~\eqref{eq:chi0}], for a linearly polarized (LP) background field ($\varepsilon=0$, black line) and a circularly polarized (CP) background field ($\varepsilon=\pm 1$, blue line). Dashed lines show the asymptotic behavior given by Eqs.~\eqref{eq:intIsmallChi} and~\eqref{eq:intIlargeChi} for the LP case. 
    The inset highlights the values of $\chi_0$ for which LP ($\chi_0 < 2.52$) or CP ($\chi_0 > 2.52$) gives higher pair production probability and the relative difference of $\mathcal{I}_{\varepsilon}$ (in absolute value) between the two cases.}
    \label{fig:fig1}
\end{figure}

Nonlinear Breit-Wheeler is the dominant QED process leading to electron-positron pair production from the collision of a high-energy photon with a strong (optical) laser pulse~\cite{Hu2010,He2020}.
As discussed in Ref.~\cite{Meuren2015}, for a relativistically intense background field, $a_0 \gg 1$, the local constant field approximation (LCFA) holds and the pair production probability can be computed by integrating the instantaneous pair production rate.
For a high-energy photon propagating with velocity ${\bf c}$,  normalized energy $\gamma_{\gamma}=\hbar\omega_{\gamma}/(m_ec^2)$ and instantaneous quantum parameter
\begin{eqnarray}
\chi_{\gamma} = \frac{\gamma_{\gamma}}{E_S}\,\sqrt{ \big({\bf E} + {\bf c} \times {\bf B}\big)^2 - \big({\bf c}\cdot{\bf E}\big)^2\!/c^2 }\,,
\end{eqnarray}
where $E_S = m_e^2c^3/(e\hbar) \simeq 1.32\times 10^{18} {\rm V/m}$ is the Schwinger field and ${\bf E}$ and ${\bf B}$ denote the electric and magnetic fields at the electron position, the instantaneous pair production rate is given by~(see, e.g., Refs.~\cite{Ritus,Baier_b_1998}):
\begin{eqnarray}\label{eq:bw-diff}
    W_{\rm\scriptscriptstyle{BW}} = W_0\,\frac{b_0(\chi_{\gamma})}{\gamma_{\gamma}}\,,
\end{eqnarray}
where $W_0 = 2\alpha^2/(3\tau_e)$.
For a fixed value of the quantum parameter $\chi_{\gamma}$, the rate is inversely proportional to the gamma photon energy $\gamma_{\gamma}$, and depends non-trivially on $\chi_{\gamma}$ through the function
\begin{eqnarray}\label{eq:b0}
b_{0} \left( \chi_{\gamma} \right) &= &\dfrac{\sqrt{3}}{2 \pi} \int_{0}^{1}  \dfrac{\rm{d} \xi}{\xi \left( 1 - \xi \right)}  \left[ \dfrac{2}{3 \chi_{\gamma}}\dfrac{1 - 2\xi}{  \left(1 - \xi\right)}\rm{K}_{5/3} \left( \mu  \right)  + \rm{K}_{2/3} \left( \mu \right) \right]\,,
\end{eqnarray}
with $\mu = 2/[3\chi_{\gamma}\xi (1-\xi)]$ and ${\rm K}_{n}(x)$ the modified Bessel function of the second kind. This function, shown in Fig.~\ref{fig:fig1}(a), can be well approximated (within less than 1\% from the numerically computed values for $\chi_{\gamma} \in [10^{-2},10^3]$) 
by\footnote{Note that $0.242 \sim 0.16 \times 3/2$, and $b_0(\chi)=3\chi\,T(\chi)/2$, with $T(\chi)$ the function used by Erber~\cite{Erber1966} and Blackburn~\cite{Blackburn2017}.}
\begin{eqnarray}\label{eq:ErberPlus}
    b_0(\chi_{\gamma}) \simeq 0.242\,\frac{{\rm K}_{1/3}^2\big(4/(3\chi_{\gamma})\big)}{1-0.172/(1+0.295\,\chi_{\gamma}^{2/3})}\, . 
\end{eqnarray}
The denominator in Eq.~\eqref{eq:ErberPlus} provides a simple improvement of the well known Erber approximation~\cite{Erber1966}, which also allows to recover the correct asymptotic behavior [dashed grey lines in Fig.~\ref{fig:fig1}(a)] of $b_0(\chi)$:
\begin{eqnarray}
    \label{eq:b0smallChi}b_0(\chi_{\gamma}) \rightarrow c_1\,\chi_{\gamma}\,e^{-8/(3\chi_{\gamma})} \quad &{\rm for}&\quad\chi_{\gamma} \ll 1\,,\\
    \label{eq:b0largeChi}b_0(\chi_{\gamma}) \rightarrow c_2\,\chi_{\gamma}^{2/3} \quad &{\rm for}&\quad\chi_{\gamma} \gg 1\,,
\end{eqnarray}
with $c_1 = (3/16)\left(3/2\right)^{3/2} \simeq 0.344$
and $c_2 = 3^{2/3}45\,\Gamma^4(2/3)/(56\pi^2) \simeq 0.569$ [$\Gamma(x)$ denoting the gamma function].

The quantity $W_{\rm\scriptscriptstyle{BW}}$ defined by Eq.~\eqref{eq:bw-diff} is the instantaneous rate of pair production, so that considering, at a time $t_0$, a given number $N_0$ of identical high-energy photons entering a strong background field, the number of photons $N_{\gamma}$ remaining at later times $t>t_0$ satisfies the rate equation:
\begin{eqnarray}\label{eq:number_photon_equa_diff}
    \frac{d}{dt}N_{\gamma} = - W_{\rm\scriptscriptstyle{BW}} N_{\gamma}\,,
\end{eqnarray}
where both $N_{\gamma}$ and $W_{\rm\scriptscriptstyle{BW}}$ depend on time (the latter when the high-energy photons explore electromagnetic fields with spatial and/or temporal variations). Note that Eq.~\eqref{eq:number_photon_equa_diff} does not account for subsequent radiation from the produced pairs, which will be verified to be accurate for the investigated parameter range in Sec.~\ref{sec::GeomEffects}. In the following $t_{0}$ will refer to the time at which the gamma flash first encounters the pulse.
The solution to this equation reads
\begin{eqnarray}\label{eq:number_photon_sol}
N_{\gamma}\left( t \right) = N_0 \, \exp\left[- \int_{t_0}^{t}W_{\rm\scriptscriptstyle{BW}}\!\left( t' \right) dt' \right]\,,
\end{eqnarray}
from which one can derive the probability $P(t_0,\Delta t) = 1-N_{\gamma}(t_0+\Delta t)/N_0$ for a given high-energy photon to decay into an electron-positron pair between time $t_0$ and $t_0+\Delta t$:
\begin{eqnarray}\label{eq:probability}
    P(t_0,\Delta t) = 1-\exp\left[- \int_{t_0}^{t_0+\Delta t}W_{\rm\scriptscriptstyle{BW}}\!\left( t' \right) dt'\right]\,.
\end{eqnarray}
When the time-integrated rate [i.e. the integral term in Eq.~\eqref{eq:number_photon_sol}] is small, it can be assimilated to the probability $P(t_0,\Delta t)$, as $P(t_0,\Delta t) \rightarrow \int_{t_0}^{t_0+\Delta t}W_{\rm\scriptscriptstyle{BW}}\!\left( t' \right) dt' \ll 1$.
This leads some authors to refer to $W_{\rm\scriptscriptstyle{BW}}$ as the pair production probability per unit time.
While this is correct in the limit ${P(t_0,\Delta t) \ll 1}$, it leads to inconsistencies as the probability increases 
and the time-integrated rate assumes values greater than unity (see, also, Refs.~\cite{Di_Piazza_2010,Meuren2015,Tamburini_2019,Podszus_2021}).
Hence, throughout this work, we will distinguish pair production rate [as defined by Eq.~\eqref{eq:bw-diff}] 
and probability [as defined by Eq.~\eqref{eq:probability}].

\subsection{Pair production probability in an arbitrarily polarized plane wave}

Let us now focus on the case of a high-energy photon colliding head-on with an ultra-intense laser pulse described by a plane wave (propagating in the $\hat{\bf z}$-direction), with electric field:
\begin{eqnarray}\label{eq:field_pw}
    {\bf E}(z,t) &=& \dfrac{E_{0}}{\sqrt{1+\varepsilon^2}}\left[\sin{ \left( \omega t - kz \right)}\,{\bf \hat{x}} + \varepsilon\cos{ \left( \omega t - kz \right)}\,{\bf \hat{y}}\right]\,,
\end{eqnarray}
magnetic field ${\bf B} = \hat{\bf z}\times {\bf E}/c$, and
polarization ellipticity $\varepsilon \in [-1,1]$. 
It takes the particular values $\varepsilon=\pm 1$ for a circularly polarized (CP) wave ($\pm$ standing for opposite helicities),
and $\varepsilon=0$ for a linearly polarized (LP) wave ($\hat{\bf x}$ denoting by convention the direction of polarization).  
It is important to stress 
that our choice of parametrisation [Eq.~\eqref{eq:field_pw}] ensures that the laser energy is the same for any value of $\varepsilon$. On the contrary, $E_0/\sqrt{1+\varepsilon^2}$ measures the maximum laser electric field amplitude and varies with the ellipticity.

When a high-energy photon collides with such a laser field (we define $t_0=0$ as the time at which the photon enters the background field), the value of its quantum parameter
\begin{eqnarray}\label{eq:chidef}
    \chi_{\gamma}(t) = \frac{\gamma_{\gamma}}{E_S}\,\Big\vert {\bf E}\big(z_{\gamma}(t),t) + {\bf c} \times {\bf B}(z_{\gamma}(t),t\big) \Big\vert\, 
\end{eqnarray}
evolves in time as the photon explores regions with different electric and magnetic field amplitude [here $z_{\gamma}(t)$ denotes the photon time-dependent position].
In the case of head-on collisions (${\bf c} = -c\,\hat{\bf z}$), the time-dependent photon quantum parameter reduces to
\begin{eqnarray}\label{eq:chi_vs_t}
    \chi_{\gamma}(t) = \chi_0\,\Psi_{\!\varepsilon}(2\omega t) 
    \quad \text{with} \quad \Psi_{\!\varepsilon}(\varphi) = \sqrt{\frac{\sin^2\!\varphi+\varepsilon^2\cos^2\!\varphi}{1+\varepsilon^2}} \,,
\end{eqnarray}
and 
\begin{eqnarray}\label{eq:chi0}
    \chi_0 = 2\,\gamma_{\gamma}\,\frac{E_0}{E_S} 
    \simeq 0.801\,\left(\frac{\hbar\omega_{\gamma}}{1\,{\rm GeV}}\right)\,\sqrt{\frac{I_0}{10^{22}~{\rm W/cm^2}}}\,.
\end{eqnarray}
Note that, with our convention on the background field parametrization [Eq.~\eqref{eq:field_pw}], $\chi_0$ denotes the maximum quantum parameter for the LP case only.
For arbitrary $\varepsilon$, the maximum achievable quantum parameter $\chi_m$ is a decreasing function of $\vert\varepsilon\vert$
\begin{eqnarray}\label{eq:chim}
    \chi_m = \frac{\chi_0}{\sqrt{1+\varepsilon^2}}.
\end{eqnarray}

Note also that Eq.~\eqref{eq:chi_vs_t} takes a simple form when considering either a CP or a LP background field:
\begin{eqnarray}
\label{eq:chi_circ}
    \chi_{\gamma}(t) = \chi_0/\sqrt{2} \quad &\text{for}& \quad \varepsilon = \pm 1\,,\\
    \chi_{\gamma}(t) = \chi_0\, \big\vert\sin(2\omega t) \big\vert \quad &\text{for}& \quad \varepsilon = 0\,.\label{eq:chi_lin}
\end{eqnarray}

\subsubsection{Decay probability after propagating through half a period of the background field\\}

For arbitrary values of $\varepsilon$, $\chi_{\gamma}(t)$ is either a constant (for $\varepsilon=\pm 1$) or a periodic function of time with period $\tau/4$ (we recall $\tau=2\pi/\omega$) corresponding to the time for the high-energy photon to cross half a wavelength of the background field. It is thus convenient to compute the probability $P_m$ of the high-energy photon to decay into a pair during this interval of time. 
Let us then consider a time $t_m$ at which the high-energy photon {\it sees} a maximum of the background electric field, and compute the probability $P_m$ for the high-energy photon to decay into an electron-positron pair in between the times $t_0=t_m-\tau/8$ and $t_0+\tau/4=t_m+\tau/8$:
\begin{eqnarray}\label{eq:Pm}
    \hspace{-2.0cm}
    P_m = P(t_0,\tau/4) = 1 - \exp\left(-R_m\frac{\tau}{4}\right) \quad \text{with} \quad 
    R_m \frac{\tau}{4} = \frac{W_0}{2\omega\gamma_{\gamma}}\int_0^{\pi}\! b_0\big(\chi_0\Psi_{\!\varepsilon}(\varphi)\big)\, d\varphi\,
\end{eqnarray}
where $R_m\tau/4$ is the time-integrate rate, while $R_m$ denotes the average rate.

Let us stress that the dimensionless quantity $W_0/(2\omega\gamma_{\gamma})$ in Eq.~\eqref{eq:Pm} is of order 1 for optical background fields (with micrometric wavelength $\lambda$) and GeV-level high-energy photons:
\begin{eqnarray}
    \frac{W_0}{2\omega\gamma_{\gamma}} = \frac{\alpha}{3}\,\frac{m_ec^2}{\hbar\omega}\,\frac{m_ec^2}{\hbar\omega_{\gamma}} 
    \simeq 0.512\,\left(\frac{\lambda}{1\,{\rm \mu m}}\right)\,\left(\frac{1\,{\rm GeV}}{\hbar\omega_{\gamma}}\right)\,.
\end{eqnarray}
The probability $P_m$, that measures the contribution of a single field maximum, depends on the integral quantity
\begin{eqnarray}\label{eq:intI}
    \mathcal{I}_{\varepsilon}(\chi_0) = \int_{0}^{\pi}\!b_0\big(\chi_0\Psi_{\!\varepsilon}(\varphi)\big)\, d\varphi\,,
\end{eqnarray}
which, for a given polarization ellipticity $\varepsilon$, is a function of $\chi_0$ only as shown in Fig. \ref{fig:fig1}(b) for LP (black line) and CP (blue line).

This integral is the key object to compute the pair creation probability and can be calculated either numerically or by means of an approximated formula as described in the following. 
In~\ref{app:ComputeIntI}, we generalize the approach proposed in~\cite{Blackburn2017} to arbitrary $\varepsilon$ and $\chi$, and show that $\mathcal{I}_{\varepsilon}(\chi_0)$ can be written in the approximate form:
\begin{eqnarray}\label{eq:intIApproximated}
    \hspace{-2.4cm}
    \mathcal{I}_{\varepsilon}(\chi_0) \simeq \pi\,b_0\!\left(\chi_m\right)\,{\rm min}\!\left\{F\big(s_{\varepsilon}(\chi_m)\big),f(\varepsilon)\right\}
    \,\text{with}\,\left\{\hspace{-1.5mm}
        \begin{array}{ll}
            F\!\left(s\right) = \sqrt{2/\pi}\,s\,{\rm erf}\!\left(\pi\sqrt{2}/(4s)\right)\\
            f(\varepsilon) = \frac{1}{\pi}\int_0^{\pi}\!\left[\sin^2\varphi+\varepsilon^2\cos^2\varphi\right]^{1/3}\!d\varphi
        \end{array}
    \right.
\end{eqnarray}
where $\chi_m$ is defined in Eq.~\eqref{eq:chim} and ${\rm erf}(x) = \tfrac{2}{\sqrt{\pi}}\int_0^x e^{-t^2}dt$ is the error function. 
The function $F(s)$ emerges from the saddle point approximation used to compute the integral when the main contribution comes from the vicinity of the field maximum. It varies slowly with $s$, as $F(s) \simeq \tfrac{2}{\sqrt{\pi}}s$ for $s\ll1$
and $F(s) \rightarrow 1$ for $s \rightarrow +\infty$.
In Eq.~\eqref{eq:intIApproximated}, $F(s)$ takes for argument $s_{\varepsilon}(\chi_m)$, with 
\begin{eqnarray}
\label{eq:sigma_vareps}
    s_{\varepsilon}(\chi) = \sqrt{\frac{3}{2}}\,\frac{c\!\left(\chi\right)}{\sqrt{1-\varepsilon^2}} 
    \quad \text{and} \quad c(\chi) = \sqrt{\frac{2\,b_0(\chi)}{3\chi\,b_0^{\prime}(\chi)}}\,,
\end{eqnarray}
where $b_0^{\prime}(\chi)$ denotes the derivative of $b_0(\chi)$ and $c(\chi)$ is a slowly varying function of $\chi$, as $c(\chi) \simeq \sqrt{\chi}/2$ for $\chi \ll 1$ and $c(\chi)\rightarrow 1$ for $\chi \rightarrow +\infty$. 

Equation~\eqref{eq:intIApproximated} is exact for CP background fields ($\varepsilon=\pm 1$) for which ${s_{\pm 1}(\chi_m)\rightarrow+\infty}$, so that $\mathcal{I}_{\pm 1}(\chi_0) = \pi\,b_0\big(\chi_0/\sqrt{2}\big)$. Moreover it allows to recover the  asymptotic behavior of Eq.~\eqref{eq:intI} in the limiting cases of small and large quantum parameter $\chi_0$:
\begin{eqnarray}
  \hspace{-2cm}\label{eq:intIsmallChi} \mathcal{I}_{\varepsilon}(\chi_0) &\xrightarrow[]{\chi_0 \ll 1}&
    \left\{
       \begin{array}{ll}
            c_3\, (1-\varepsilon^2)^{-1/2}(1+\varepsilon^2)^{-1/4}\,\chi_0^{3/2}\,
            \exp\left(-\frac{8\sqrt{1+\varepsilon^2}}{3\chi_0}\right) \quad {\rm for} \quad \chi_0 \ll 1-\varepsilon^2\,, \\
           c_4\, \chi_0\,\exp(-\frac{8\sqrt{2}}{3\chi_0}) \quad {\rm otherwise}\,, 
       \end{array}
    \right.\\
   \hspace{-2cm}\label{eq:intIlargeChi}\mathcal{I}_{\varepsilon}(\chi_0) &\xrightarrow[]{\chi_0 \gg 1}&c_5\,\frac{f(\varepsilon)}{(1+\varepsilon^2)^{1/3}}\,\chi_0^{2/3}\,,
\end{eqnarray}
where we used Eqs.~\eqref{eq:b0smallChi} and~\eqref{eq:b0largeChi}, and $c_3 = 27\sqrt{\pi/2}/64 \simeq 0.529$, $c_4 = 9\sqrt{3}\pi/64 \simeq 0.765$ and ${c_5 = 3^{2/3}45} \Gamma^4(2/3)/(56\pi)$ ${\simeq 1.789}$.

\subsubsection{Total decay probability\\}\label{sec:totalDecayProb}

As will be further discussed in Sec.~\ref{sec:theoryDiscussion}, considering parameters typical of the upcoming generation of laser facilities, the probability for a high-energy photon to decay into a pair after crossing a single half-wavelength of the background field will in general be small compared to 1. This will not however be the case when considering the cumulative effect of crossing several wavelengths. 

The probability for a high-energy photon to decay into a pair after $t=n\tau/4$ of interaction with the background field (or, equivalently, after crossing $n$ local maxima of the field amplitude) $P_{\rm tot}(t=n\tau/4)$, can be derived from the number $N_{\gamma}$ of photons surviving after a time $t$: 
\begin{eqnarray}
    \hspace{-1cm}
    N_{\gamma}(t=n\tau/4) = N_0\,\prod_{m=1}^n \widebar{P}_m \quad \text{with} \quad \widebar{P}_m = 1-P_m = \exp(-R_m\tau/4)\,.
\end{eqnarray}
Here we have used $m$ as a running index for compactness, and $P_m$ [given by Eq.~\eqref{eq:Pm}] and $\widebar{P}_m$ denote the probabilities for the photon to decay
and {\it not to} decay, respectively, during the $m^{\rm th}$ interval. 
We can then write the total probability of producing a pair as
\begin{eqnarray}\label{eq:Ptot}
    \hspace{-1cm}
    P_{\rm tot}(t=n\tau/4) = 1 - \prod_{m=1}^n \widebar{P}_{m} = 1 - \exp\left(-R\,t\right)\quad \text{with} \quad R = \frac{1}{n}\sum_{m=1}^n R_m\,,
\end{eqnarray}
$R$ denoting the average pair production rate.

\begin{figure}\centering%%%
    \includegraphics[width=0.9\textwidth]{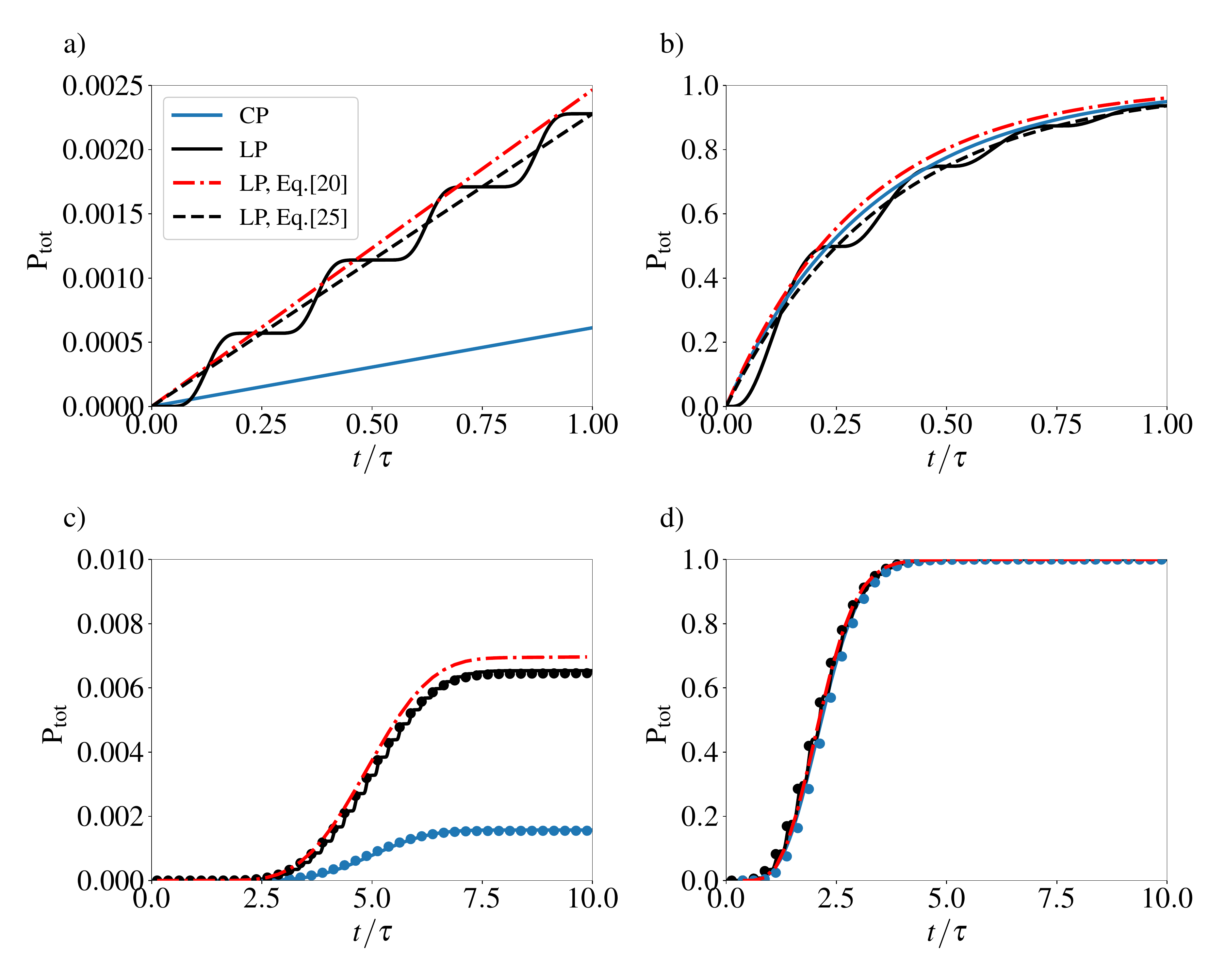}
    \caption{
    Temporal evolution of the probability for a high-energy photon with normalized energy $\gamma_{\gamma}=\hbar\omega_{\gamma}/(m_ec^2)=10^3$
    to decay into an electron-positron pair while interacting with a plane wave background field in head-on collision.
    Considering a constant background field envelop with (a) $a_0\simeq 82$ 
    corresponding to $\chi_0 = 0.5$ and (b) $a_0\simeq 660$ corresponding to $\chi_0 = 4$.
    Considering a background field with sin$^2$ temporal profile with maximum field amplitude
    (c) $a_0\simeq 82$ corresponding to $\chi_0 = 0.5$, (d) $a_0\simeq 660$ corresponding to $\chi_0 = 4$.
    Solid lines correspond to the {\it exact} probability obtained by integrating numerically Eq.~\eqref{eq:probability}, dashed lines and dots to the theoretical prediction obtained from Eq.~\eqref{eq:Ptot}, for LP ($\varepsilon=0$, black) and CP ($\varepsilon=\pm 1$, blue) cases. The red dash-dotted lines correspond to the theoretical prediction from Eq.~\eqref{eq:Ptot} using the approximation of Eq.~\eqref{eq:intIApproximated} for LP.}
    \label{fig:probVStime}
\end{figure}

Let us note that this formula is exact at $t=n\tau/4$ with $n \in \mathbb{N}$, and -- as will be shown later -- can also be used with good approximation at all times $t \gg \tau/4$. Moreover, in the case of a monochromatic plane wave (no temporal envelop), the average pair production rate simply reduces to $R=R_m$. When dealing with a pulse with a finite temporal envelop, however, Eq.~\eqref{eq:Ptot} needs to be computed combining the contribution of successive $(\tau/4)$-long intervals, using the local maximum of the background field strength for each time interval. This approach gives very good results even considering ultra-short, few cycles laser pulses. To confirm this, Figure~\ref{fig:probVStime} shows the temporal evolution of the decay probability for a high-energy photon with $\gamma_{\gamma} = 10^3$ colliding head-on with different background fields. 
In  panels (a) and (b), the high-energy photon interacts with a background field with constant amplitude $a_0\simeq 82$ [$\chi_0=0.5$, Fig.~\ref{fig:probVStime}(a)] and $a_0\simeq 660$ [$\chi_0=4$, Fig.~\ref{fig:probVStime}(b)]. The interaction lasts for a time $\tau$, i.e. the photon explores two wavelengths of the background field.
In panels (c) and (d), the high-energy photon interacts with a background field with a sin$^2$ temporal profile in intensity, a full-width-half-maximum ({\sc fwhm}) of $5\tau$, and maximum field strength $a_0\simeq 82$ and $a_0\simeq 660$, respectively.

In all panels, the solid lines denote the {\it exact} probability computed by numerically integrating Eq.~\eqref{eq:probability}, considering either LP (black lines) or CP (blue lines) background field. We can see that, for $\chi_{0}= 0.5$, CP produces less pairs than LP, while for $\chi_{0}= 4 $ CP is slightly more efficient [the difference between the two polarizations will be discussed in details in Sec.~\ref{sec:theoryDiscussionA}].
These probabilities are compared against the prediction of Eq.~\eqref{eq:Ptot}, computed using either the numerically evaluated $\mathcal{I}_{\varepsilon}(\chi_0)$ [dashed lines in panels (a) and (b), dots in panels (c) and (d)] or the approximation given by Eq.~\eqref{eq:intIApproximated} to compute $R_m$ (red dot-dashed lines, only computed for LP as it is exact for CP). Note that here and in the following, whenever Eq.~\eqref{eq:intIApproximated} is used, the improved Erber approximation Eq.~\eqref{eq:ErberPlus} is exploited.

Panels (a) and (b) show an excellent agreement between the exact computations (solid lines) and Eq.~\eqref{eq:Ptot} (dashed lines), both using the numerically integrated values of $\mathcal{I}_{\varepsilon}(\chi_0)$.
A remarkably good agreement is also found when considering the fully analytical approximation of Eq.~\eqref{eq:intIApproximated} for $\mathcal{I}_{\varepsilon}(\chi_0)$ (red dot-dashed lines). The probability is slightly overestimated in this case, as expected for $\chi_0 \sim 1$ for which Eq.~\eqref{eq:intIApproximated} shows the greatest departure from the exact expression of $\mathcal{I}_{\varepsilon}(\chi_0)$.

Similarly, an excellent agreement between all three approaches is found in panels (c) and (d) considering a short ($5\tau$ {\sc fwhm} in intensity) background field. In this case the value of $R_m$ in each $\tau/4$ interval is computed by taking the value of $\mathcal{I}_{\varepsilon}$ given by the local maximum of the field, i.e. the time variation of the envelope is considered slow during $\tau/4$. The validity of this assumption is confirmed by PIC simulations discussed in the following. 
In particular, Figure~\ref{fig:probVStime} shows that the approach leading to the derivation of Eq.~\eqref{eq:Ptot}, which consists in treating the cumulative contributions of successive maxima of the background field, remains a very good approach even for ultra-short, few cycle, background fields.

Last, we note that the results of 1D PIC simulations performed with \Smilei (see~\ref{app:1Dsimulations} for details) are not distinguishable from the {\it exact} computations (solid lines) for all reported cases, and are therefore not shown in Fig.~\ref{fig:probVStime}.

\subsection{Guidelines for pair production optimizations}\label{sec:theoryDiscussion}

\subsubsection{Influence of the background field polarization\\}\label{sec:theoryDiscussionA}

Interesting insights into the role of the background field polarization on pair production
can be gained from the asymptotic behavior provided by Eqs.~\eqref{eq:intIsmallChi} and~\eqref{eq:intIlargeChi}.
 
We recall that, motivated by experimental constraints, we compare LP and CP at fixed energy: this condition implies that the CP beam has a lower maximum amplitude. Because of this, for $\chi_0 \ll 1$, Eq.~\eqref{eq:intIsmallChi} shows that any departure from the LP case leads to a tremendous decrease of $\mathcal{I}_{\varepsilon}(\chi_0)$.
Indeed the exponential cut-off of the pair production rate is strongly affected by the decrease of $\chi_m$  with $\vert\varepsilon\vert$ as shown in Eq.~\eqref{eq:chim}.
In contrast, for $\chi_0 \gg 1$, $f(\varepsilon)/(1+\varepsilon^2)^{1/3}$ is an increasing function of $\vert\varepsilon\vert$: the reduction of the field peak amplitude is compensated by the fact that the absolute value of the amplitude is not time dependent [see Eq.\eqref{eq:chi_circ}].
Hence, in this range, increasing the background field ellipticity increases the pair production rate.
However, the rate increase from the LP to the CP case is small (less than $12\%$), and it is thus of marginal use to consider CP for boosting pair production\footnote{In particular as, in practice, implementing CP on multi-petawatt facilities would lead to a decrease of the delivered laser pulse energy.}. 

The differences between the LP and CP cases are clearly shown in the inset in Fig.~\ref{fig:fig1}(b), where $\mathcal{I}_{0}(\chi_0)$ is found to be orders of magnitude larger than $\mathcal{I}_{\pm 1}(\chi_0)$ for $\chi_0 \ll 1$ while $\mathcal{I}_{\pm 1}(\chi_0)$ is only about 10\% larger than $\mathcal{I}_{0}(\chi_0)$ for $\chi_0 \gg 1$. 
Similar conclusions can be drawn from Fig.~\ref{fig:probVStime}. For small values of the photon parameter, $\chi_0 = 0.5$ (left panels), the pair production probability is significantly larger considering a LP background field instead of a CP one. For the higher value of $\chi_0 = 4$ (right panels), CP increases, but only marginally, pair production.
For this reason, we will from now on focus on LP background fields, unless otherwise specified.

\subsubsection{Influence of the background field amplitude and gamma-photon energy\\}\label{sec:theoryDiscussionB}
\begin{figure}\centering
    \includegraphics[width=0.9\textwidth]{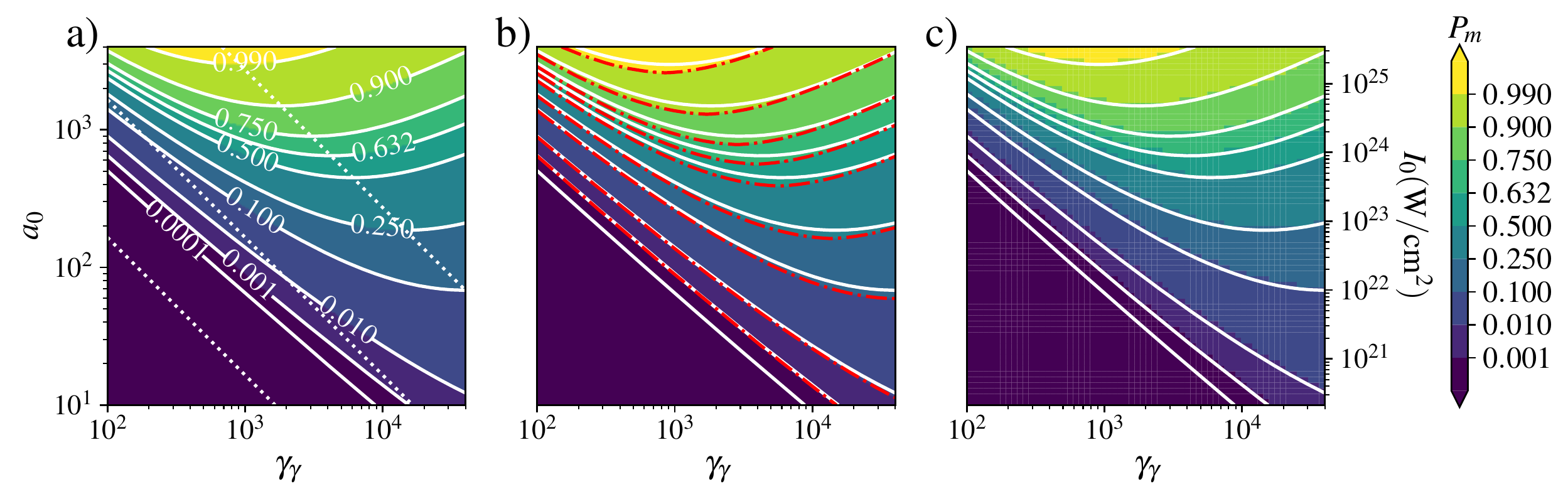}
    \caption{
        Probability $P_m$ 
        for a high-energy photon to decay into an electron-positron pair 
        after crossing half-a-wavelength of a LP background field, 
        as a function of the photon energy $\gamma_{\gamma}=\hbar\omega_{\gamma}/(m_ec^2)$ and background field amplitude $a_0$:
        (a) integrating numerically Eq.~\eqref{eq:Pm}, 
        (b) computed using the approximate expression Eq.~\eqref{eq:intIApproximated},
        {(c)~extracted} from one-dimensional PIC simulations.
        In all three panels, the solid white lines report the isocontours of the first panel. In panel b), the isocontours obtained from Eq.~\eqref{eq:intIApproximated} are shown in red dot-dashed lines. White dotted lines correspond to constant values of $\chi_{0}= 0.1, 1, 16.5$ from bottom left to top right corners.}
    \label{fig:Pm}
\end{figure}

The relative influence on pair production of the photon energy $\gamma_{\gamma}$ and the background field amplitude $a_0$ (or equivalently intensity $I_0$, as we consider here a given background field wavelength $\lambda=0.8\, {\rm \mu m}$), is summarized in Fig.~\ref{fig:Pm}. 
It shows the probability $P_m$ for a high-energy photon to decay in a pair after crossing half a wavelength of a LP background field:
(a) integrating numerically Eq.~\eqref{eq:Pm} which provides an {\it exact} value of $P_m$;
(b) using the approximate but fully analytical expression given by Eq.~\eqref{eq:intIApproximated};
{(c)~extracted from} 1D PIC simulations of the interaction of a flash of high-energy photons colliding head-on with a plane wave (details are given in \ref{app:1Dsimulations}).

Let us first discuss panel (a) and note that, in addition to the probability isocontours (solid white lines), the dotted white lines represent the contours of constant quantum parameter, $\chi_0=0.1$, $1$ and $16.5$. As $\chi_0 \propto \gamma_{\gamma}\,a_0$, these are straight lines with a $-45^{\circ}$ inclination. 

In the limit $\chi_0 \ll 1$, i.e. the bottom-left of Fig.~\ref{fig:Pm}, $P_m$ assumes very small values. In this limit, the time-integrated rate $R_m\tau/4$ scales as $\gamma_{\gamma}^{-1}\chi_0^{3/2}\exp\!\left[-8/(3\chi_0)\right]$ [Eq.~\eqref{eq:intIsmallChi}] and the exponential term, depending only on $\chi_0$, gives the dominating contribution to $P_m$. As a result, the isocontours of $P_m$ (solid white lines) behave as nearly straight lines, roughly parallel to the contours of constant $\chi_0$. 
We wish to stress however that the dependence on $\gamma_{\gamma}^{-1}$ cannot be fully ignored: the isocontours of $P_m$ are less steep than the isocontours of $\chi_0$ which indicates that $P_m$ increases faster with $a_0$ than with $\gamma_{\gamma}$. 
In this range of small $\chi_0$, for which probability and time-integrated rate are equivalent, the importance of increasing $a_0$ to improve pair production was already pointed out in~\cite{Blackburn2017}. However, the probability variation with both $a_0$ and $\gamma_{\gamma}$ needs to be examined in more details at higher $\chi_0$.

Indeed, the dependence with $\gamma_{\gamma}^{-1}$ plays a more important role as $\chi_0$ increases 
so that $\chi_0$ can not be considered as the only relevant parameter in order to optimize the pair production rate. 
This is clearly seen by considering the isocontours of $P_m$ in ($a_0$,$\gamma_{\gamma}$)-plane,  Fig.~\ref{fig:Pm}(a): the isocontour slope becomes shallow and eventually changes sign. Thus, for any value of $P_m$, a minimum field strength $a_0$ is needed in order to obtain the desired level of probability.
A corollary is that, at constant $a_0$, increasing $\gamma_{\gamma}$ increases the probability $P_m$ up to a maximum value, beyond which a further increase of $\gamma_{\gamma}$ would only decrease $P_m$. The minimum of each isocontour can thus be found by solving $\partial P_m/\partial\gamma_{\gamma} = 0$, which one can recast in the form
\begin{eqnarray}\label{eq:turningPoint}
\dfrac{\rm{d}\mathcal{I}_{\varepsilon}}{\rm{d}\chi_{0}} - \dfrac{\mathcal{I}_{\varepsilon}(\chi_{0})}{\chi_{0}} = 0\,.
\end{eqnarray}
Interestingly, this equation involves only $\chi_{0}$ so that the minima of the probability isocontours lie on a straight line of constant $\chi_{0}$. This {\it a priori} surprising result can be better understood noting that both $a_0$ and $P_m$ are Lorentz invariants, the minimum value of $a_0$ to reach a given probability $P_m$ can depend only on the Lorentz invariant involving the photon energy, i.e. $\chi_0$. Solving numerically Eq.~\eqref{eq:turningPoint} for $\varepsilon = 0$ (LP case), we obtain\footnote{Solving Eq.~\eqref{eq:turningPoint} for arbitrary values of $\varepsilon \in [-1,1]$ leads to $\chi_0$ in between $16.5$ and $17.2$, and corresponding values of $I_{\varepsilon}(\chi_0)$ in between $5.3$ and $6.0$.} $\chi_{0} \simeq 16.5$, which is reported by a dotted line in Fig. \ref{fig:Pm}.

To conclude with Fig.~\ref{fig:Pm}, let us now turn to panels (b) and (c).
Panel (b) reports the probability $P_m$ computed from Eq.~\eqref{eq:intIApproximated} and the corresponding isocontours (red dashed lines). Superimposing the isocontours (white lines) of the probability from panel (a), we can see an excellent agreement between the two approaches. This validates the approximate form, Eq.~\eqref{eq:intIApproximated}, which has the advantage to be completely analytical and can now be used to get quick yet precise estimates of the probability $P_{m}$. 
Last, panel (c) reports the probability extracted from 1D PIC simulations of the head-on collision of a flash of high-energy photons with a laser beam. 
The probability was extracted from the depletion of high-energy photons after crossing half a laser wavelength. Here again, a very good agreement is observed with the isocontours (white lines) obtained from the {\it exact} integration shown in panel (a). This last panel thus provides a cross-benchmark of our model and the Monte-Carlo module for Breit-Wheeler pair production implemented in our PIC code \Smilei.\\

\begin{figure}\centering
    \includegraphics[width=0.6\textwidth]{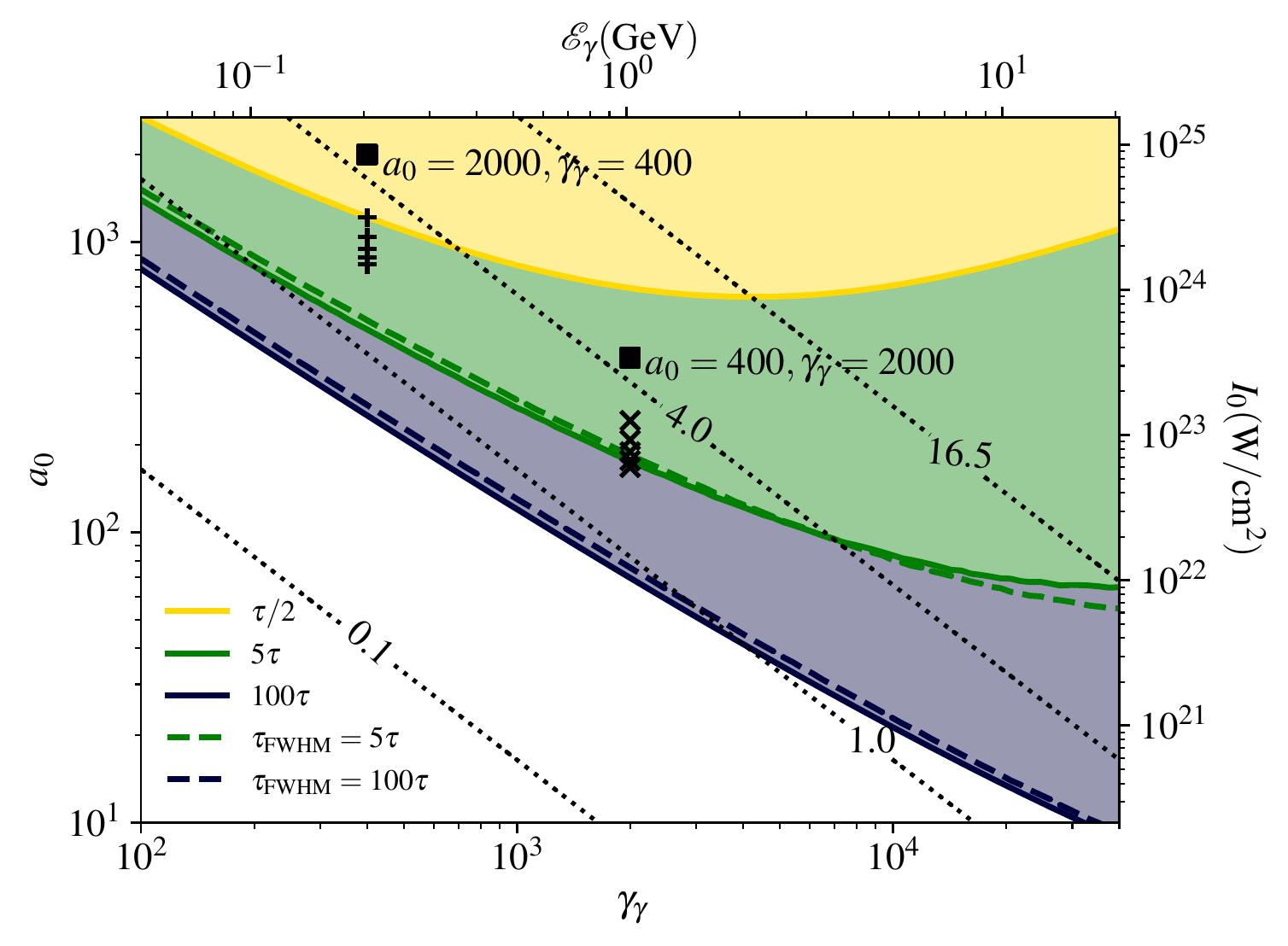}
    \caption{
    Position in the ($\gamma_{\gamma},a_0$)-plane [equivalently in the ($\mathcal{E}_{\gamma},I_{0}$)-plane] at which $R \Delta t = 1$ [$P_{\rm tot}(\Delta t) \simeq 0.63$] for the interaction with a single maximum of the incoming pulse (half-wavelength, i.e. pulse duration equal to $\tau/2$) for which the interaction lasts for $\Delta t= \tau/4$ (yellow line) and for the interaction with pulses with a step-like time profile (solid lines) of duration $5\tau$ (green) and $100\tau$ (blue), and pulses with a $\sin^{2}$ time envelope with {\sc fwhm} of $5\tau$ (green dashed line) and $100\tau$ (blue dashed line). The color-shaded areas above the plain lines correspond to $R \Delta t > 1$ (considering the step-like time profiles). The dotted black lines correspond to constant $\chi_{0}=0.1\,, 1\,, 4\,, 16.5\,$. The symbols highlight the parameters of the simulations discussed in Sec.~\ref{sec:sim1} and~\ref{sec:sim2}.}
    \label{fig:fig4}
\end{figure}

Our model can be used to predict the effectiveness of pair production for a given set of laser and high-energy photon parameters\footnote{In the case of large secondary pair production, our prediction will underestimate the number of pairs as our analysis is limited to the soft-shower regime.}.  
Most of the photons crossing the laser will be converted in pairs over a time $\Delta t$ if the quantity $R \Delta t$ in Eq.~\eqref{eq:Ptot} becomes of order~1, for which\footnote{Note that in this limit, the probability cannot be assimilated to the time-integrated rate, as in the case of very weak pair production.} $P_{\rm tot}(\Delta t) \geq 0.63$. 
In Fig.~\ref{fig:fig4}, we highlight in yellow the region in the $(a_{0},\gamma_{\gamma})$-plane where the probability for a high-energy photon to decay into a pair after interacting with a single maximum (half-wavelength) of the laser pulse $P_{\rm tot}(\tau/4)$ is equal or larger than $0.63$ (the pulse duration is here $\tau/2$, but the time required for the photon to explore half the laser wavelength is $\Delta t = \tau/4$).
The condition $P_{\rm tot}(\Delta t) = 0.63$ is also shown for pulses with a step-like time profile (solid lines) and duration $5\,\tau$ (green, $\Delta t=2.5\tau$) or $100\,\tau$ (blue, $\Delta t=50\tau$), and pulses with a $\sin^{2}$ intensity time profile with {\sc fwhm} $5\tau$ (green dashed line, $\Delta t = 5\tau$) and $100\,\tau$ (blue dashed line, $\Delta t = 100\tau$). Note that both step-like and $\sin^2$ time profiles correspond to laser pulses with the same energy and peak power, and lead to similar requirements to reach $P_{\rm tot} \simeq 0.63$. 
Comparing the blue and green lines, we see that at constant $\gamma_\gamma$ a longer pulse satisfies the condition of efficient conversion for a lower value of $a_{0}$.
Hence, even though the condition $R\,\tau/4 = 1$ (yellow solid line) is quite stringent and achieving a high probability level over a time interval $\tau/4$ can be difficult, the condition $R \,\Delta t \gtrsim 1$ is significantly relaxed considering longer interaction time. Indeed, significant (order~1) pair production probability is expected on forthcoming multi-PW laser facilities (see Sec.~\ref{sec:discussion} for details).

Let us finally note that, when $R\,\tau/4 \geq 1$ (yellow area), more than $63\%$ of conversion is achieved after crossing a single wavelength of the background field, and more than $99\%$ of the incident gamma photons are converted into pairs in less than 3 periods. This gives a very robust condition for abundant pair creation, and it determines a limit above which is not useful to further increase the laser amplitude to increase the number of primary pairs. 
As we will show in the following, this condition ($R\,\tau/4 \geq 1$) can also be invoked to find the optimal laser transverse shape and focal spot for a given laser energy. Indeed 3D PIC simulations, shown by the symbols in Fig.~\ref{fig:fig4} and discussed in detail in the next Sec.~\ref{sec::GeomEffects}, show a different behaviour depending on whether their initial parameters lie in the yellow or green region of Fig.~\ref{fig:fig4}. In the case of the two simulations reported as squares in the figure, we will see that these differences are observed even though the simulation parameters correspond to the same value of $\chi_0$.

\section{Geometrical effects on pair production: the case of LG beams}\label{sec::GeomEffects}

This Section aims to generalize the model developed in Sec.~\ref{sec:totalDecayProb}, to overcome the plane wave approximation and account for the laser beam transverse profile.
We start by reminding some properties of the LG beams, limiting ourselves to the critical aspects for the understanding of pair production.
We then compare the predictions of our theoretical model (generalized to account for the laser transverse profile in Sec.~\ref{sec:modelAppliedToLG}) with the results of 3D PIC simulations, whose initialization is discussed in Sec.~\ref{sec:simulationSetup}. 
A thorough comparison of the efficiency of pair production considering beams with the same energy and different field structures is presented in Secs.~\ref{sec:sim1} and \ref{sec:sim2}. We explore first the impact of the beam geometry and amplitude in a regime of efficient pair production (time-integrated rate $R_m \tau/4 >1$) and then in a regime ($R_m \tau/4 <1$) more relevant for current experimental facilities (as will be discussed in the following Sec.\ref{sec:discussion}). Finally in Sec.~\ref{sec:PairPropagation}, we discuss some properties of the produced pairs, such as the energy spectrum and the propagation direction.  

\subsection{Properties of LG beams}\label{sec:LGbeamProperties}
The LG modes are solutions of the paraxial equation with cylindrical symmetry and a generalization of the Gaussian beam. 
Each mode is indexed by two integers: $p \geq 0$ for the radial order and $\ell$ for the azimuthal one. Hence, the notation ${\rm LG}_{p\ell}$ will be used in the following to identify a specific mode. 

Starting from Eq.~\eqref{eq:field_pw}, the electric field of an LG beam can be expressed using $E_0= a_0\,  \mathfrak{R}\left(u_{p\ell }\right)$, where the complex envelope $u_{p\ell }$ in cylindrical coordinates is~\cite{siegman} 
\begin{eqnarray}
\nonumber
    u_{p\ell } \left( \rho, \phi, z \right) &=& C_{p\ell }\,\dfrac{ w_{0} }{w(z)}\left( \dfrac{\sqrt{2} \rho}{w(z)} \right)^{|\ell |} L_{p}^{|\ell |}\!\left(\dfrac{2\rho^{2}}{w^2(z)} \right)\\
\label{eq:LG_field}    
    &\times& \exp{\left[- \dfrac{\rho^{2}}{w^2(z)}\right]} \exp{\left[ -i\psi_{pl} \left( z \right) + i\ell \phi + i \dfrac{z \rho^{2}}{w^2(z)}\right]}\, ,
\end{eqnarray}
where $C_{p\ell } = \sqrt{ p! /(p+|\ell |)!} $ is a normalization constant, $L_{p}^{\ell}(x)$ are the generalized Laguerre polynomials, $w(z) = w_{0}\sqrt{1 - z^2/z_R^2}$ is the beam waist with $w_{0} = w(z = 0)$ the waist at focus and $z_{R} = \pi w_{0}^2/\lambda$ the Rayleigh length, and $\psi_{p\ell }\left( z \right) = \left( 2p + \vert \ell \vert +1  \right) \arctan{\left( z/z_{R} \right)} $ is the generalized Gouy phase. 
Note that the LG modes described by Eq.~\eqref{eq:LG_field} have all the same energy and that ${\rm LG}_{00}$ corresponds to a Gaussian beam with maximum field amplitude $a_0$. \\

\begin{figure}\centering
    \includegraphics[width=0.9\textwidth]{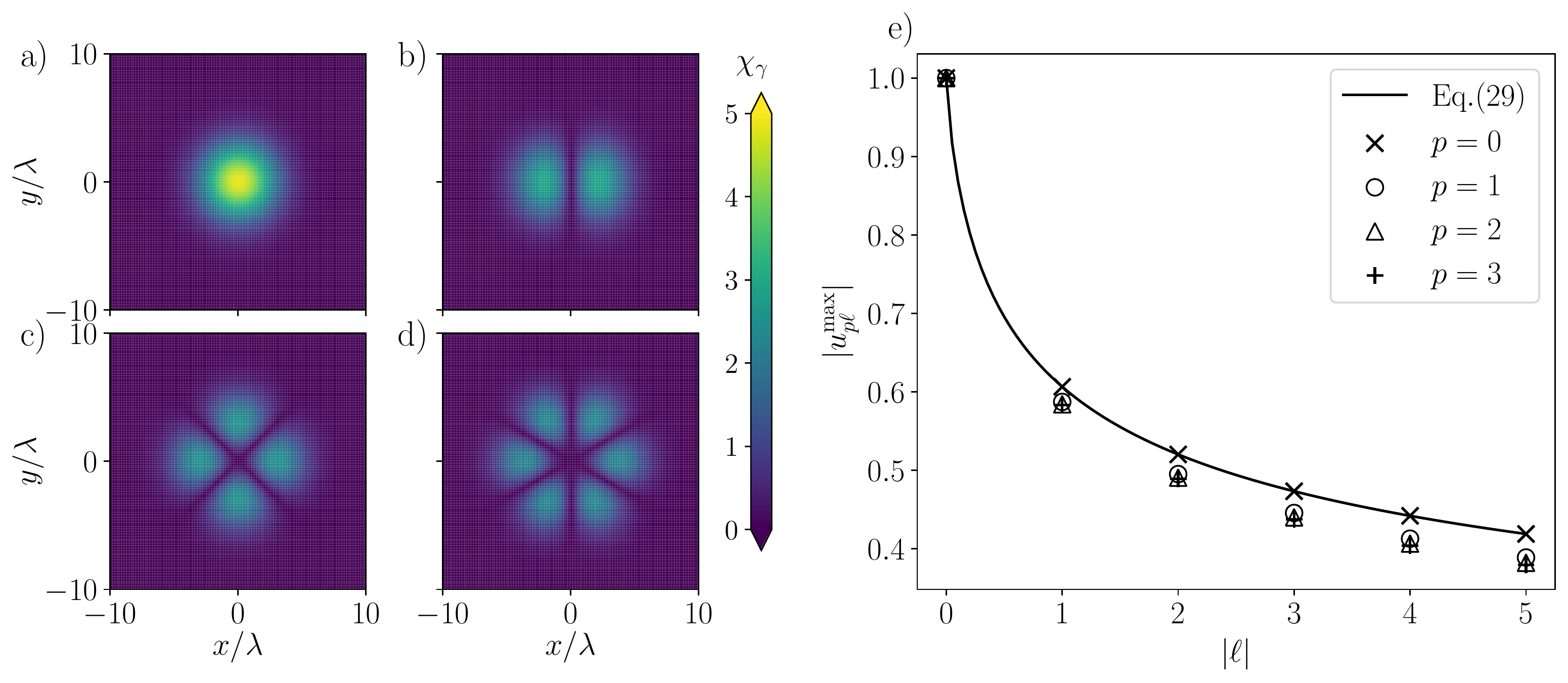}
     \caption{Photon quantum parameter $\chi_{\gamma}$ in the head-on collision of photons of energy $\gamma_{\gamma} = 400$ with a LP LG beam at focus [i.e. $z=0$ in Eq.~\eqref{eq:LG_field}] with $a_0 = 2000$, $p = 0$ and a) $\ell = 0$, b) $\ell  = 1$, c) $\ell  = 2$, d) $\ell  = 3$. e) Maximum of the complex envelope $u^{\rm max}_{p\ell}$ from Eq.~\eqref{eq:umax} (black line) and for $p \in  [0,3]$ as a function of $|\ell|$. For $p \neq 0$, Eq.~\eqref{eq:drhou0} is solved numerically.}
      \label{fig:LG_beams}
\end{figure}

In order to infer how the  structure of the LG modes might affect pair production we show in Fig.~\ref{fig:LG_beams}(a)-(d) the quantum photon parameter $\chi_{\gamma}$ for the head-on collision of photons of energy $\gamma_{\gamma} = 400$ with a LP LG beam at focus (i.e. $z=0$) with $a_0 = 2000$, $p = 0$ and $\ell \in  [0,3]$.
As can be seen in Fig.~\ref{fig:LG_beams}(a)-(d), for $\ell\neq 0$ the maximum of the field, and thus the peak values of the quantum parameter, are no longer along the beam axis (i.e. $x=y=0$), in contrast with the Gaussian beam case. The radius at which the field has its maximum amplitude $\rho^{\rm max}$ can be obtained by solving
$\partial_{\rho} |u_{p\ell}|^2 = 0$, leading to:
\begin{eqnarray}\label{eq:drhou0}
 \frac{2 \rho^2}{w(z)^2}
+\left( 1-\delta_{0p} \right) \frac{4 \rho^2}{w(z)^2} \frac{L_{p-1}^{|\ell| + 1}\!\left( \frac{2\rho^2}{w(z)^2} \right)}{L_{p}^{|\ell|}\!\left( \frac{2\rho^2}{w(z)^2} \right)} 
- \vert\ell\vert
 = 0 \, ,
\end{eqnarray}
with $\delta_{ij}$ the Kronecker delta.
For $p = 0$, Eq.~\eqref{eq:drhou0} can be solved analytically and gives $\rho^{\rm max} = w(z) \sqrt{|\ell|/2}$.
While the radial position of the field maximum for $p=0$ increases with $|\ell|$, the amplitude decreases, as shown in Fig.~\ref{fig:LG_beams}(e), and tends asymptotically to zero for $|\ell| \gg 1$ as $ |2 \pi \ell|^{-1/4}$. The maximum amplitude of the envelope $u_{0\ell }$ can be computed inserting $\rho^{\rm max}$ into Eq.~\eqref{eq:LG_field}, which gives
\begin{eqnarray}\label{eq:umax}
u^{\rm max}_{0\ell}=|u_{0\ell}(\rho^{\rm{max}},\varphi,z)| = \frac{1}{\sqrt{1 + z^{2}/z_{R}^{2}}} \, \frac{|\ell|^{|\ell|/2}\, {\rm e}^{-|\ell|/2}}{\sqrt{|\ell|!}}  \, .
\end{eqnarray}
Figure~\ref{fig:LG_beams}(e) also shows $u^{\rm max}_{p\ell}$ for $p \neq 0$, obtained by inserting the numerical solution of Eq.~\eqref{eq:drhou0} into Eq.~\eqref{eq:LG_field}. The maximum field amplitude for the LG beams decreases with both $|\ell|$ and $p$, and as a consequence the same behaviour is observed for the maximum value of $\chi_{\gamma}$ [see Fig.~\ref{fig:LG_beams}(a)-(d)]. 
Therefore, in the following we will consider only $|\ell| \leq 5$ and $p = 0$, because of the higher field amplitude (hence higher values of $\chi_{\gamma}$) and because of the practical challenges in producing high order modes at relativistic intensity.  

Note that in the considered cases with $\ell \neq 0 $, $\chi_{\gamma}$ has $2|\ell|$ maxima in the transverse plane [see Fig.~\ref{fig:LG_beams}(b)-(d)] and, looking at the evolution in time, each maximum makes a full rotation around the laser propagation axis in $|\ell| \tau$. This means that, in a fixed transverse plane, a maximum rotates to the location of the next one in half a period. 

If we consider the head-on collision of the LG beam with a gamma photon, the photon is crossing half a period of the field in $\tau/4$. Hence, the photon crosses a rotating local maximum in $\tau/4$. This observation is fundamental for generalization of the reduced model presented in the previous Sec.~\ref{sec:totalDecayProb}, where we consider subsequent intervals of duration $\tau/4$ to infer the total probability.

As noted in Fig.~\ref{fig:LG_beams}, the LG beams not only have a non-trivial spatial distribution of the fields in the transverse plane, but also a maximum field amplitude decreasing with $|\ell|$ [following Eq.~\eqref{eq:umax}]. In order to distinguish the impact on pair production of these two aspects, we discuss the interaction with extended Gaussian (EG) beams. For each LG beam (i.e. each value of $|\ell|$), we define the associated EG beam as the Gaussian beam that has the same maximum field amplitude and carries the same total energy. As $u^{\rm max}_{p\ell}$ decreases with $|\ell|$ [Eq.~\eqref{eq:umax}] and the energy is kept constant, the waist of the EG beams increases with respect to the fundamental Gaussian (equivalent to $\rm LG_{00}$) as $w_{\ell} =w_{0} \sqrt{|\ell|!}\,{\rm e}^{|\ell|/2} |\ell|^{-|\ell|/2} \geq w_{0}\, .$ 

\subsection{Simulation set-up}\label{sec:simulationSetup}

\begin{figure}\centering
    \includegraphics[width=0.5\textwidth]{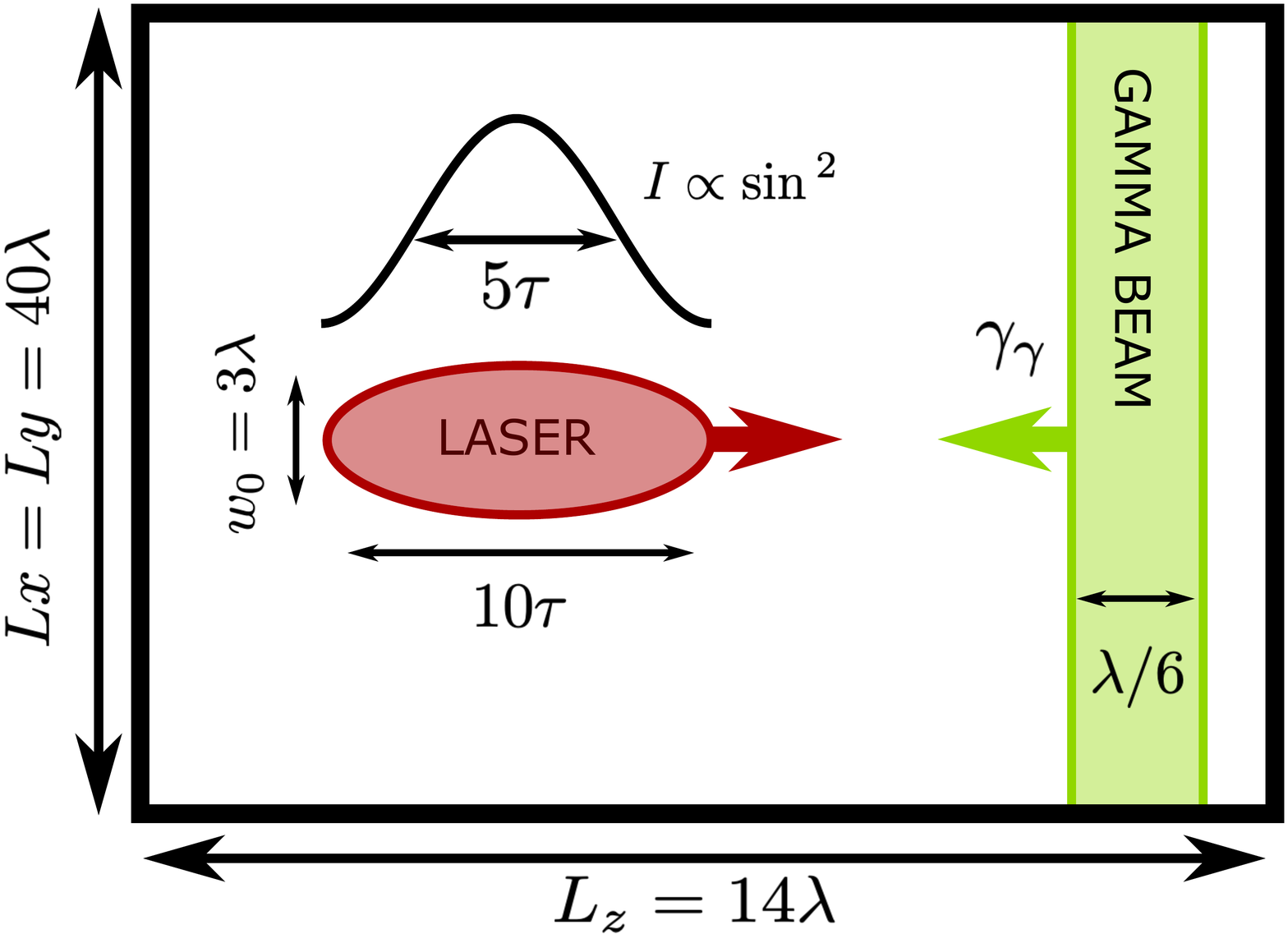}
      \caption{Typical set-up of the 3D PIC simulations, with $\lambda$ and $\tau$ the laser wavelength and the optical cycle, respectively.}
      \label{fig:setup_simu}
\end{figure}

In what follows, we present 3D PIC simulations performed with the open-source code \Smilei~\cite{smilei}, considering the setup schematically represented in Fig.~\ref{fig:setup_simu}. 
Each simulation reproduces a volume of $40\lambda\times 40\lambda\times 14\lambda$ (in the $x$, $y$ and $z$ directions, respectively) with spatial resolution $\lambda/24$ and temporal resolution at 95\% of the CFL condition~\cite{Nuter2014}. 

An intense laser pulse (with wavelength $\lambda = 0.8\,\mu \rm m$) is injected from the left boundary $z=-5\lambda$ using the method presented in Ref.~\cite{Perez2019} that allows for the Maxwell-consistent injection of an electromagnetic wave with an arbitrary spatio-temporal profile.
Here, this method is used to inject the laser pulse by prescribing its spatio-temporal profile in its focal plane [i.e. the $(x,y)$-plane at $z=0$], as given by the real part of Eq.~\eqref{eq:LG_field} at $z=0$, multiplied by a temporal envelope so that the laser pulse has a $\sin^2$ temporal profile in intensity with {\sc fwhm} $\tauFWHM = 5\,\tau$, and total duration of $10\, \tau$. 
The laser beam collides head-on with a counter-propagating flash of high-energy photons with a finite longitudinal extension $L_\gamma = \lambda/6$, transverse extension equal to the simulation box, and a density $n_{\gamma}$ equal to the critical density $n_c = \epsilon_0 m_e \omega^2/e^2$. 
The duration of the simulation is long enough for the two light beams to stop interacting, but short enough so that none of the photons or secondary particles escape from the simulation box. All results presented hereafter have been extracted at the end of the simulations.

In Secs.~\ref{sec:sim1} and \ref{sec:sim2}, we present two series of simulations: the first one with $a_0=2000$ and $\gamma_{\gamma}=400$, the second with $a_0=400$ and $\gamma_{\gamma}=2000$, both series corresponding to a reference quantum parameter $\chi_0 \simeq 6 \times 10^{-6}\, a_0\,\gamma_{\gamma}  \sim 4.85$ (for $\lambda=0.8~{\rm \mu m}$). For each series, we varied the transverse spatial profile of the high-intensity laser pulse in order to study the impact of the spatio-temporal field profile, comparing different LG beams (varying $\ell$, but keeping $p=0$) and the corresponding EG beams.  
Note however that within each series, $a_0$ is left unchanged and refers to the maximum field amplitude of the reference Gaussian beam (for which we consider $w_{0} = 3 \lambda$). As a result, within each set of simulations, the laser pulse energy is unchanged, equal to $\simeq \rm 10\,kJ$ for $a_0=2000$ and $\simeq \rm 410\,J$ for $a_0=400$. Conversely, the photon flash surface energy density $m_e c^2\,\gamma_{\gamma} n_\gamma d$ is $\simeq 5\,\rm mJ/\lambda^2$ for the first series (with $\gamma_{\gamma}=400$) and $\simeq 25\,\rm mJ/\lambda^2$ for the second one ($\gamma_{\gamma}=2000$).

As highlighted in Fig.~\ref{fig:fig4} (squares), the reference Gaussian case of the first series of simulations allows us to explore the physics in the regime of high pair production probability after a single $\tau/4$ interval (i.e. $R_m \tau/4 >1$, yellow region). On the contrary, with the second series we investigate a regime where efficient pair production is achieved thanks to the cumulative effects of several wavelength (i.e. $R_m \tau/4 <1$, but the reference Gaussian case is still above the green line in Fig.~\ref{fig:fig4}).

\subsection{Reduced model for pair creation with arbitrary spatio-temporal laser beam profiles}\label{sec:modelAppliedToLG}

In this Section we generalize the model developed in Sec.~\ref{sec:totalDecayProb} to accurately describe pair production from complex, spatially structured laser beams. 

To do so, we exploit the fact that we are here dealing with high-energy photons colliding head-on with the strong background field. Hence, the photon momenta are left unchanged as the photon crosses the background field, and their trajectories are straight lines so that the field seen by a given photon along its trajectory is known and depends only on the photon initial position ($x_0, y_0$) in the plane transverse to the direction of propagation of both light beams\footnote{For analytical tractability, we consider all photons to be propagating exactly in the $z$-direction.}. It follows that the quantum parameter of any photon is known throughout the photon interaction with the background field, and for a given background field, is determined uniquely by the photon initial position ($x_0, y_0$). As a result, the total probability $P_{\rm tot}$ for any photon to decay into a pair after interacting with the laser pulse is also determined by the photon initial position, and the total number of produced pairs at the end of the interaction between the flash of high-energy photons (with density $n_{\gamma}$ and longitudinal width $L_\gamma$) simply reads
\begin{eqnarray}\label{eq:sigmatot}
N_{pair} = n_{\gamma} L_{\gamma}\,\sigma_{\rm tot} \quad \text{with} \quad \sigma_{\rm tot} = \int\!dx_0\,dy_0\,P_{\rm tot}(x_0,y_0)\,.
\end{eqnarray}
A key quantity to model pair production is thus the {\it total cross-section} $\sigma_{\rm tot}$, here defined as the integral of the total probability over the transverse plane.

To compute the total probability at any position ($x_0, y_0$) we use Eq.~\eqref{eq:Ptot} and proceed as described in Sec.~\ref{sec:totalDecayProb}, reconstructing the total probability from successive time intervals of duration $\tau/4$ (as discussed in Sec.~\ref{sec:LGbeamProperties}, a photon at any position in the transverse plane explores in $\tau/4$ a local maximum of the field amplitude whatever value $\ell$ assumes). One then only needs to compute the maximum field amplitude satisfactorily. For simplicity, we do so measuring the field amplitude at focus, taking the absolute value of the field envelope from Eq.~\eqref{eq:LG_field} at $z=0$. This approach, which neglects the effects of diffraction, is justified whenever the duration of the interaction ($\sim\tauFWHM$) does not exceed the time $z_R/c$ for light to cross the laser pulse Rayleigh length. As shown in the following, this approximation proves satisfactory considering typical ultra-short (Ti:Sapphire) light pulses.

\begin{figure}
  \centering
    \includegraphics[width=0.7\textwidth]{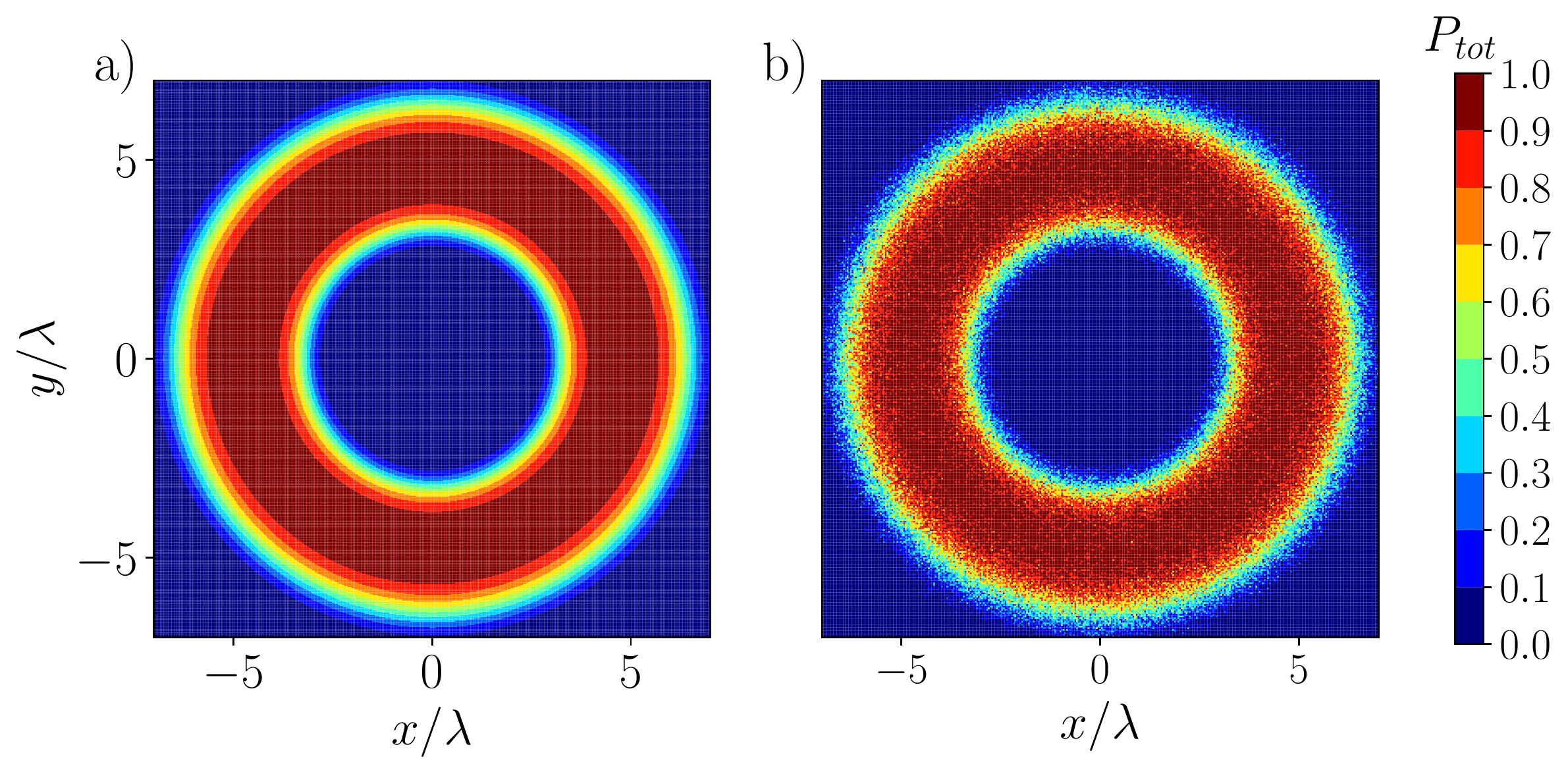}
      \caption{Probability map for a photon with energy $\gamma_{\gamma} = 400$ to produce a pair during the interaction with the $\rm LG_{05}$ beam with {\sc fwhm} of $5\tau$ and maximum amplitude of the reference Gaussian beam equal to $a_{0} = 2000$, corresponding to $\chi_{0} \simeq 4.85$. a) Analytical prediction from Eq.~\eqref{eq:Ptot}. b) Numerical results from the 3D PIC simulation.}
      \label{fig:proba_map}
\end{figure}

The result of this procedure is shown in Fig.~\ref{fig:proba_map}(a) for the case of a LG beam with $\ell=5$, $p=0$ and $a_0=2000$ colliding with a flash of photons with energy $\gamma_\gamma=400$ (corresponding to $\chi_0=4.85$; all other parameters are specified in Sec.~\ref{sec:simulationSetup}). 
Panel (a) shows a map, in the transverse plane, of the probability $P_{\rm tot}$ computed over the whole interaction time in the transverse plane. Integrating over the transverse plane provides the measure of the total cross-section $\sigma_{\rm tot} \simeq 91.4 \,\lambda^2$.

To test the impact of the approximations used in the model (phase and diffraction effects are neglected), we show in Fig.~\ref{fig:proba_map}(b) the probability map extracted from a 3D PIC simulation, where all effects neglected in our model are taken into account. Here the probability is extracted at the end of the interaction as the ratio of the remaining photon linear density by the initial one.
An excellent  agreement is found with the theoretical prediction of panel (b), with a measured total cross-section $\sigma_{\rm tot} \simeq 91.0\,\lambda^2$, which confirms the predictive capability of our simple model for the case of spatially structured beams.

\subsection{Simulations in the regime of high pair production probability $(R_{m} \tau/4 >1)$}\label{sec:sim1}

\begin{figure}
    \centering
    \includegraphics[width=0.9\linewidth]{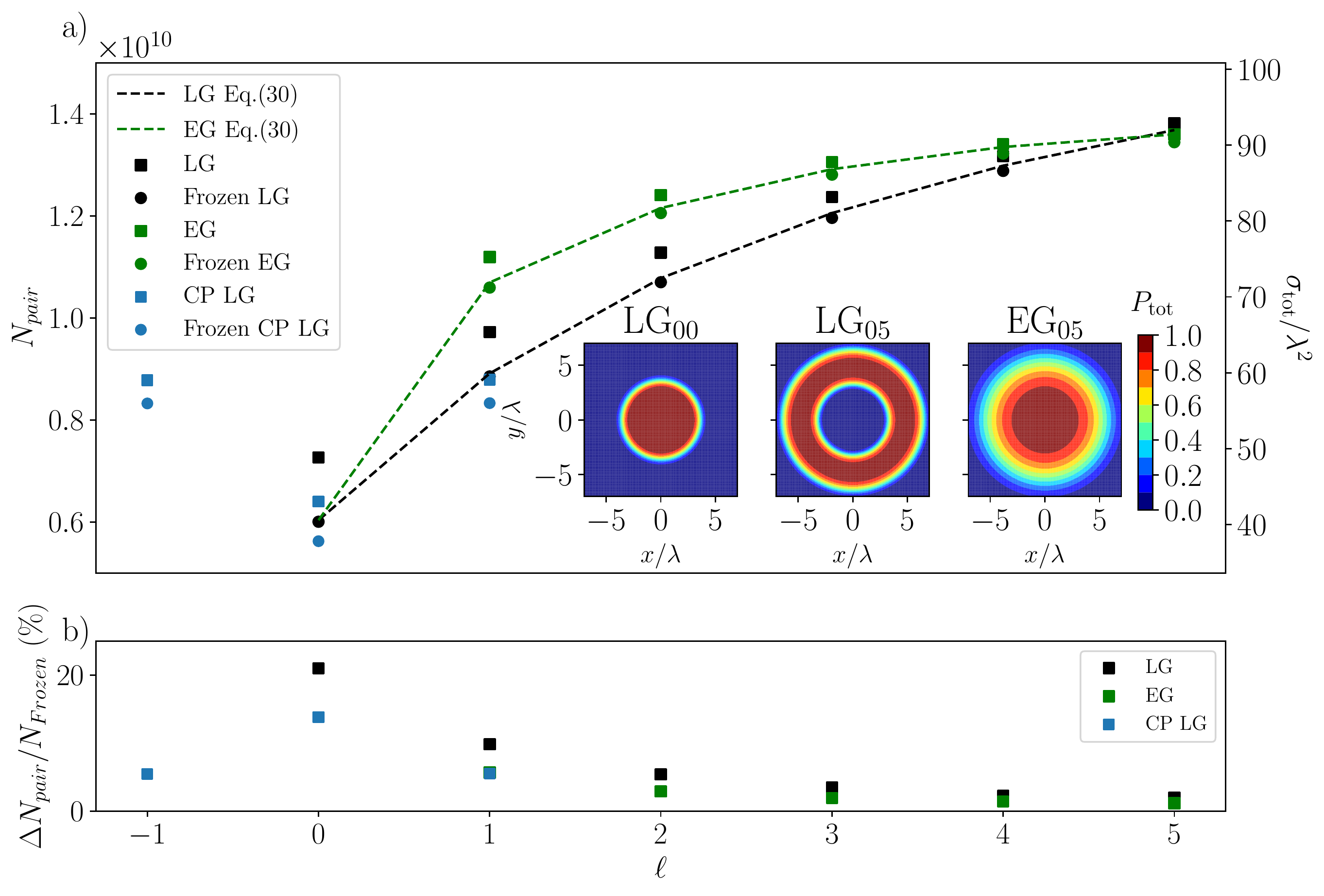}
    \caption{a) Number of produced pairs, and corresponding cross section $\sigma_{\rm tot}/\lambda^{2}$ after the interaction of gamma photons with energy $\gamma_{\gamma} = 400$ and a counter-propagating LG with $p=0$, $\ell \in [\![-1,5]\!]$ and $a_0 = 2000$ [CP (blue), LP (black)], and the corresponding LP EG (green). Frozen simulations (dots), full simulations (squares) and theoretical predictions from Eq.~\eqref{eq:sigmatot} (dashed lines). The insets show the probability map for the $\rm LG_{00}$, $\rm LG_{05}$ and $\rm EG_{05}$ cases. b) Relative difference in the number of produced pairs between the frozen and full simulation results.}
    \label{fig:sim1}
\end{figure} 

In order to investigate the dependence on the laser structure of the number of produced pairs, we present here a study in the regime satisfying $R_m \tau/4 >1$ for the reference Gaussian beam, in which we expect efficient pair production. We remind that we consider here photons with normalized energy $\gamma_{\gamma} = 400$ and a reference Gaussian beam with maximum amplitude $a_{0} = 2000$, leading to $\chi_{0} \simeq 4.85$ (all other parameters have been specified in Sec.~\ref{sec:simulationSetup}). 

To test our analytical predictions and the assumptions at the base of our model, we performed simulations (hereafter referred to as {\it frozen cases}) in which the produced pairs are prevented to further radiate. In this condition, no secondary pairs are produced and the simulations allow to directly test the proposed analytical model. We also performed simulations including the full system dynamics (hereafter referred to as \textit{full cases}), meaning that the produced pairs interact with the laser fields and potentially radiate. This allowed us to verify that we are still in the soft-shower regime, and that only a small departure from predicted number of produced pairs is indeed observed.

Figure~\ref{fig:sim1}(a) shows the number of produced pairs at the end of the interaction for all the tested cases, LP LG (in black), CP LG (blue) and EG (green), dots refer to the frozen simulations and squares to the full ones.
Considering a LP LG laser beam, pair production efficiency increases with $\ell$.
This means that for the high value of $a_0$ considered here, it is preferable to increase $\ell$, even if it is decreasing the maximum amplitude [as shown in Fig.~\ref{fig:LG_beams}(e)], as it increases the characteristic transverse size of the beam, and so the interaction area where the probability is high [see inset in Fig.~\ref{fig:sim1}(a)]. 
The EG beams (which have by construction the same maximum amplitude of the LG beams but Gaussian profile) are more efficient than the corresponding LG, up to to $\ell=4$, as they have a bigger region with substantial probability in their probability maps. However, a slightly greater number of pairs are produced in the simulation with $\rm LG_{05}$ than $\rm EG_{05}$ beam. This because the maximum field amplitude has dropped substantially and a large part of the interaction region in the EG case has low probability of pair production [see inset in Fig.~\ref{fig:sim1}(a)]. 
Predictions form our model discussed in Sec.~\ref{sec:modelAppliedToLG} and based on Eq.~\eqref{eq:sigmatot}, shown in dashed lines, are in very good agreement with both EG and LG simulations.

Based on the prediction from Fig.~\ref{fig:fig1}, for the considered $\chi_0\simeq 4.85$ we would expect at first look a higher efficiency in the CP case than with LP, contrary to what is obtained from PIC simulations [blue symbols in Fig.~\ref{fig:sim1}(a)]. However, given the spatial structure of the laser field in the transverse plane, in a large part of the interaction region the quantum number is $\chi<2.5$, i.e. the value above which CP should be favoured based on the analysis shown in Fig.~\ref{fig:fig1}(b) obtained within the plane wave approximation. Hence, the slightly higher number of pairs produced in the CP case with respect to LP in the region where $\chi>2.5$, cannot compensate the higher efficiency of LP in the region of $\chi<2.5$.  
Note that the results using a CP beam are independent from the sign of $\ell$ (i.e from the total angular momentum). This suggests that the total angular momentum of the laser has no effect on the number of pair produced. 
%For completeness we should mention we used unpolarised cross sections in the simulations, however, according to \cite{King2013} this should not change the results for our configurations. %However, since simulations were done using unpolarized cross sections for the pair production, the result has to be taken carefully.

For completeness, we compare the results of the frozen cases with simulations reproducing the full dynamics of the system [squares in Fig.~\ref{fig:sim1}(a)]. 
As highlighted in Fig.~\ref{fig:sim1}(b), the difference in the produced number of pairs is always below $25\%$ of its value in the frozen case, and it decreases with $|\ell|$. This confirms that the majority of pairs comes from the conversion of primary gamma photons and not from subsequent radiation, as expected for a soft shower. 
In this regime, our model, which accurately captures the physics of pair production for the frozen simulations (dashed lines), can be therefore used to predict pair production in realistic conditions. 
\begin{figure}
    \centering
    \includegraphics[width=0.9\linewidth]{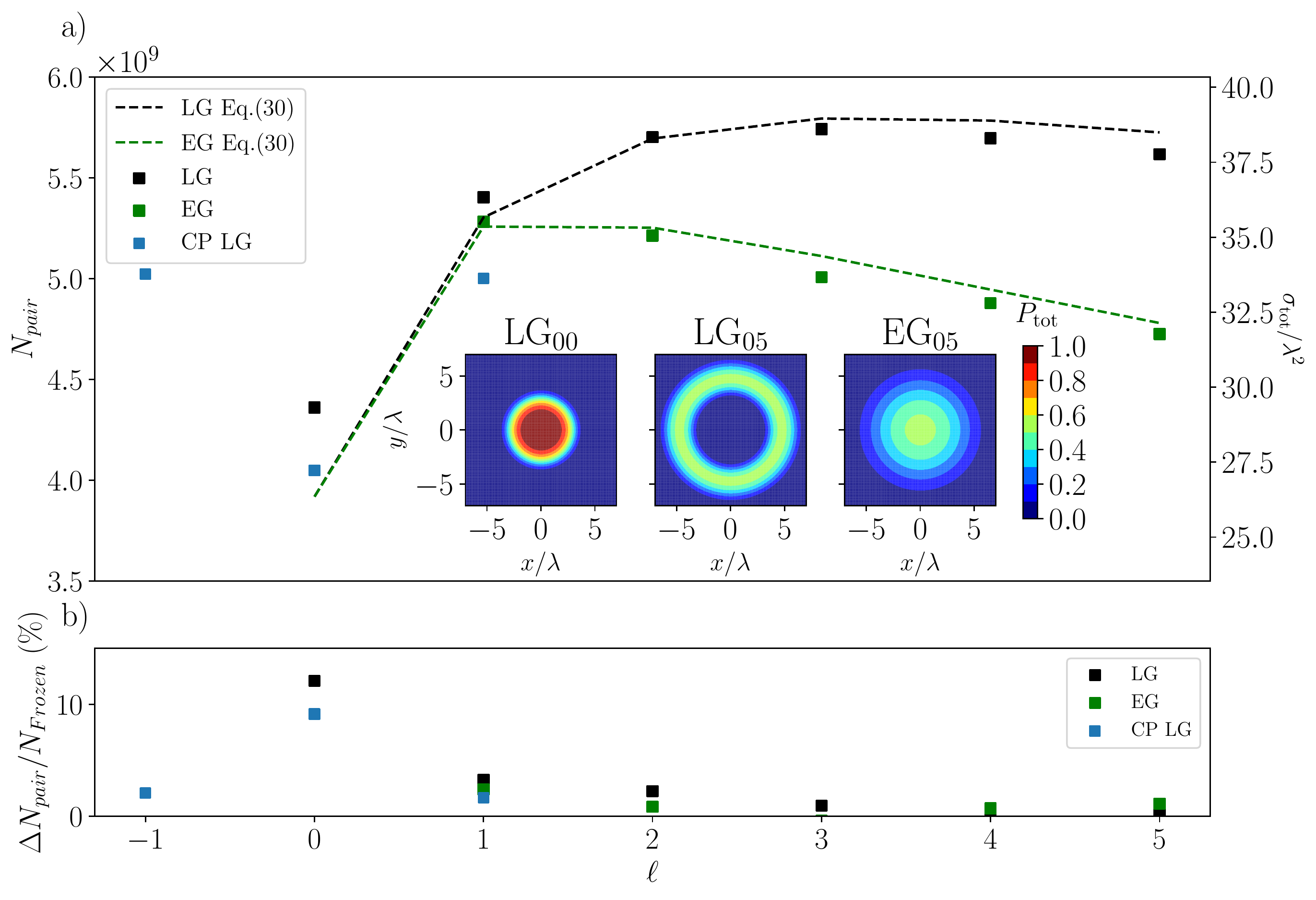}
    \caption{a) Number of produced pairs and corresponding cross section $\sigma_{\rm tot}/\lambda^{2}$, after the interaction of gamma photons of energy $\gamma_{\gamma} = 2000$ with a counter-propagating LG beam with $p=0$, $\ell \in [\![-1,5]\!]$ and $a_0 = 400$ [CP (blue) and LP (black)], and the corresponding LP EG (green). Full simulations (squares) and theoretical predictions from Eq.~\eqref{eq:sigmatot} (dashed lines). The insets show the probability map for the $\rm LG_{00}$, $\rm LG_{05}$ and $\rm EG_{05}$ cases. b) Relative difference in the number of produced pairs between the frozen and full simulation results. The frozen cases are not shown in panel (a) as the difference with the full ones is always below 15\%.}
    \label{fig:fig9}
\end{figure}

\subsection{Simulations in the regime of moderate pair production probability $(R_{m} \tau/4 <1)$}\label{sec:sim2}

In this second series of simulations we explore the regime of $R_m \tau/4 <1$, maintaining the same maximum $\chi_0$ as in the previous section by exchanging the values of field amplitude (now equal to $a_{0} = 400$) and photons energy (now equal to $\gamma_{\gamma} = 2000$). 
Note that considering a laser duration ($\tauFWHM=5\tau$), we still expect efficient pair production (the reference plane wave simulation lies above the green dashed line in Fig.~\ref{fig:fig4}). This can be seen in the inset of Fig.~\ref{fig:fig9}(a) for the Gaussian case $\rm LG_{00}$ where a significant region exhibits a probability of order $1$.

As Figure~\ref{fig:fig9} shows, for the simulations with LG beams the number of produced pairs is optimized for $\ell = 3$. The increase for $|\ell| \leq 3$ can be explained as in the previous section by geometrical considerations, i.e. the area with the field being large enough to reach $P_{\rm tot}$ of order one is getting wider with $|\ell|$. In the contrary for $|\ell| > 3$, the laser field amplitude becomes too low and having a laser beam with a larger transverse extension does not improve pair production. This is due to the fact that the field is becoming too low, so that the effect of decreasing $P_{tot}$ on pair production is more important, and is not compensated anymore by the increase in transverse size.

Moreover, in this regime, all LG beams perform better than the corresponding EG, given that the region with high (even though smaller than in the previous Sec.~\ref{sec:sim1}) probability is larger for the LG than EG, as shown in the inset of Fig.~\ref{fig:fig9} for $\ell=5$. 
Also, a similar behaviour as discussed in the previous Sec.~\ref{sec:sim1} is observed for the CP simulations and can be explained with the same reasoning. Last but not least, our model is reproducing with high accuracy the observed number of pair in this case too.

\subsection{Energy and angular distribution of the produced pairs}\label{sec:PairPropagation}
\begin{figure}
    \centering
    \includegraphics[width=0.9\linewidth]{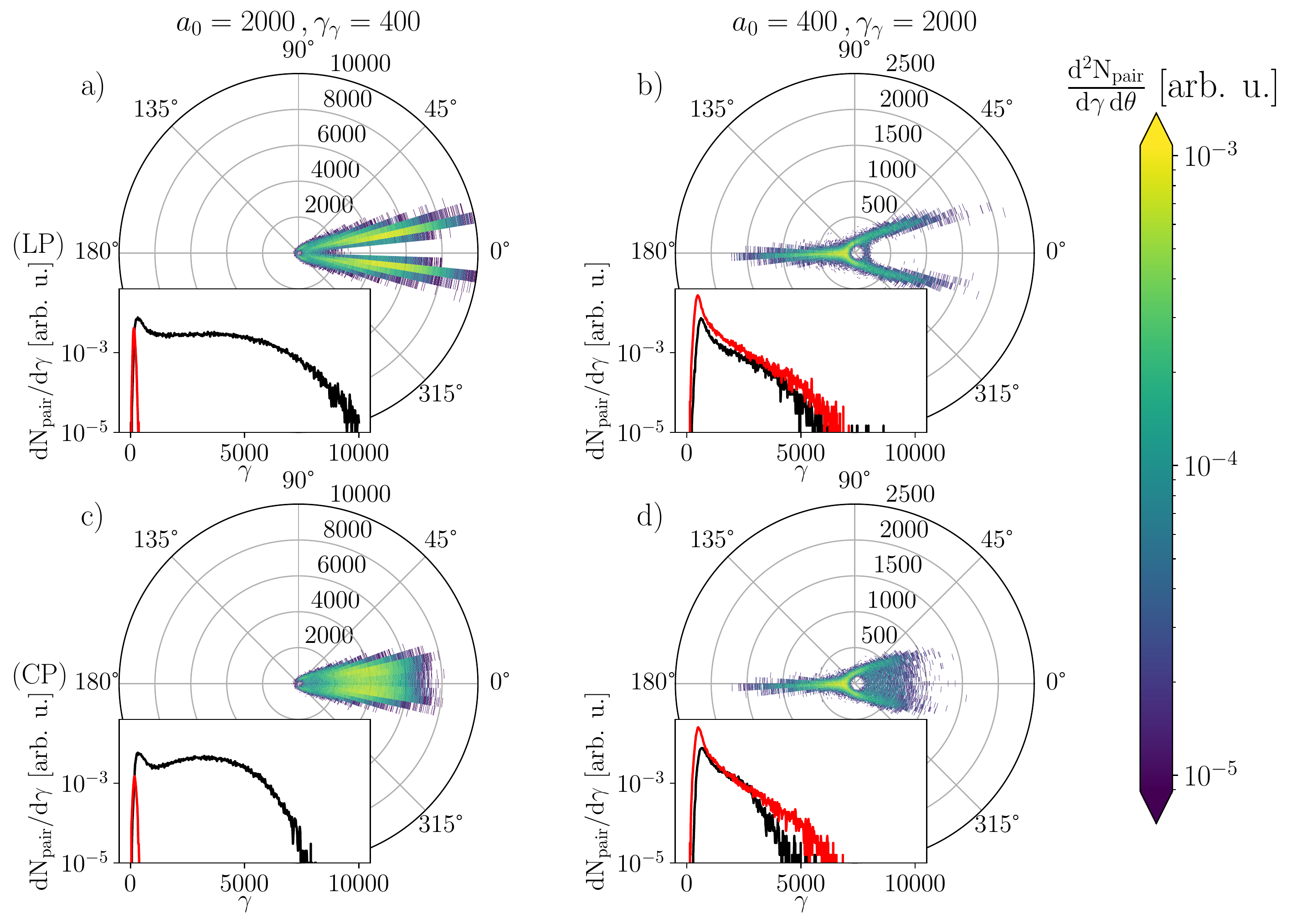}
    \caption{Positron distribution in Lorentz factor $\gamma$ (equiv. normalized energy) and angle $\theta=\arctan(p_x/p_z)$ for the first simulation series (left column) and for the second one (right column); LP (CP) in the top (bottom) row. The insets show the energy spectra for positrons propagating in the positive (black line) and in the negative (red line) directions.}
    \label{fig:fig10}
\end{figure}

The energy-angular distribution of the produced pairs in the $z-x$ plane (i.e the plane formed by the laser propagation direction $\hat{\bf z}$ and the polarization direction $\hat{\bf x}$ for the LP case) recorded at the end of the interaction is shown in Fig.~\ref{fig:fig10} considering the reference LP and CP Gaussian beams. The top row gives the positron distribution in $\theta=\arctan(p_x/p_z)$ and in energy for the LP beams and the bottom one for the CP beams. The insets show the energy spectra of particles moving in the positive (same as the laser, black line) and negative (same as the gamma flash, red line) $z$-directions.

At high $a_0$ (panels a and c) the produced pairs, initially generated with momentum in the negative $z$-direction, are predominantly moving in the same direction as the laser (corresponding to $\theta=0^{\circ}$) within a cone of aperture $40^{\circ}$, meaning that they have been slowed down and then pushed by the laser. They reach a maximum energy of $\simeq 10^4\, m_ec^2$ for the LP case and $\simeq 8 \times 10^3\, m_ec^2$ for the CP beam, consistent with the decrease of the peak field amplitude in CP with respect to LP at constant energy.

The spatial distribution is drastically modified when considering $a_0=400$. Even if there still is a substantial amount of pairs propagating in the laser direction, the majority of them keep their original direction, along the initial gamma photons propagation axis and anti-parallel to the laser pulse one. For these pairs the maximum energy is independent from the polarization, as shown in Fig.~\ref{fig:fig10}(b)-(d), contrarily to the pairs that are slowed down and pushed by the laser (right side of the quadrants or black line in the insets). 

This study, supported by complementary simulations (not shown), suggests that for $a_0\gtrsim \gamma_{\gamma}$ most of the created particles will end up propagating along the laser direction, while for smaller values of the laser maximum amplitude a large fraction of pairs will keep their original direction. A complete characterisation of the pairs spectrum and directionality is beyond the scope of this work, and will be investigated in more details in the future, as it can give important information for the planning of upcoming experimental campaigns.

\section{Discussion}\label{sec:discussion}

\begin{figure}
    \centering
    \includegraphics[width=0.9\linewidth]{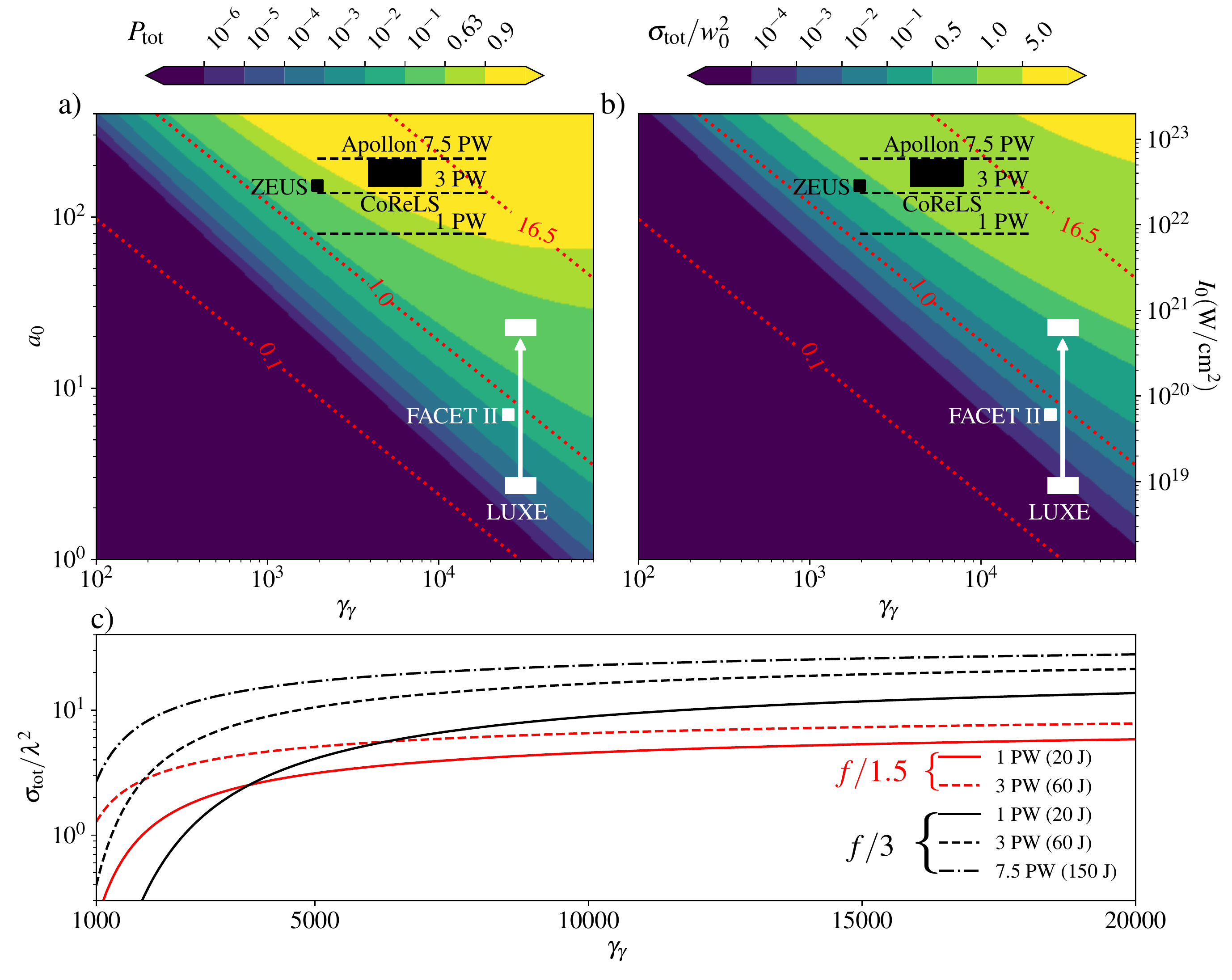}
    \caption{Model predictions considering a laser pulse with $\sin^2$ time-envelope in intensity ($\tau_{\rm{FWHM}} = 25\, \rm{fs}$) and $\lambda = 0.8\,\mu \rm{m}$. (a) Maximum pair production probability from Eq.~\eqref{eq:Ptot}. (b) Total cross section normalised to $w_{0}^{2}$ using the model discussed in Sec.~\ref{sec:modelAppliedToLG}. In both panels (a) and (b) the red lines denote contours of constant $\chi_{0} = 0.1 \, , \, 1 \, , \, 16.5$. The regimes accessible with current and upcoming facilities are also reported.
    c) Total cross section normalised to $\lambda^2$ as a function of $\gamma_{\gamma}$ computed for the Apollon facility operating at 1 PW (solid lines), 3 PW (dashed lines) and 7.5 PW (dash-dotted line) and with focusing aperture $f/3$ (black lines) and $f/1.5$ (red lines). In this last panel, at a given power, the laser energy is the same for the different focusing aperture $f/N$.}
    \label{fig:fig11}
\end{figure}

The model developed in this work allows to predict the pair production capability of upcoming facilities. In Fig.~\ref{fig:fig11}, we discuss the pair production maximum probability and total cross-section as a function of the photon energy and laser amplitude. The considered parameter range is relevant to ultra-short Ti:Sapphire ($\lambda = 0.8~{\rm \mu m}$) lasers, where studies of high-field physics and QED are already planned~\cite{Lee2018,Meuren2020,Zhang2020,Grech2021}. Two types of facilities can be distinguished.
First, the Apollon\footnote{The Apollon, located on the {\it Plateau de Saclay}, south of Paris, France, started operating at the PW-level~\cite{Apollon,Papadopoulos2016}; see also the \href{https://apollonlaserfacility.cnrs.fr/en/home}{Apollon website}.}, 
CoReLS\footnote{CoReLS, in Gwanju, South Korea, is operating at up to 4PW~\cite{corels}; see also the \href{https://corels.ibs.re.kr/html/corels_en}{CoReLS website}.} and {\sc zeus}\footnote{A 3PW laser facility being built at the University of Michigan; see the \href{https://zeus.engin.umich.edu}{{\sc zeus} website}.} facilities (shown in black in Fig.~\ref{fig:fig11}) are designed to deliver multiple light beams with duration in between 15 and 30 fs ({\sc fwhm}) and peak power from 1 to 10 PW. 
Second, the {\sc facet-ii}\footnote{{\sc facet-ii} is planned at SLAC National Accelerator Facility~\cite{Yakimenko2019}; see also the \href{https://facet.slac.stanford.edu/overview}{{\sc facet-ii} website}.} and {\sc luxe}\footnote{{\sc luxe} is an experiment proposed at the European XFEL~\cite{abramowicz2019}; see also the \href{https://luxe.desy.de}{{\sc luxe} experiment website}.} experiments will couple conventional electron accelerators with high-intensity (100 TW-class) ultra-short laser pulses (in white in Fig.~\ref{fig:fig11}).

In Fig.~\ref{fig:fig11}(a,b) the accessible $a_0$ and typical $\gamma_\gamma$ envisioned with these facilities (following the reference articles) are reported. For Apollon only, we computed $a_0$ considering a 20~fs light pulse delivering 20, 60 and 150~J on target, corresponding to peak power of 1, 3 and 7.5~PW, respectively. The reported value of $a_0$ is then computed for a LP Gaussian beam with focusing aperture $f/3$ (where the beam waist for a focusing aperture $f/N$ is $w_0 \simeq 0.90\,N\lambda \simeq 2.2~{\rm \mu m}$), using
\begin{eqnarray}
a_{0} & \simeq& \dfrac{151}{N}\sqrt{\dfrac{\mathcal{E}_{\rm laser}}{10 \rm{J}}} \sqrt{\dfrac{25\, \rm{fs}}{\tauFWHM}} \simeq \dfrac{239}{N} \sqrt{\dfrac{P_{\rm laser}}{1 \, \rm{ PW}}} \,.
\end{eqnarray}
Note that, as most of these facilities will operate in the $\chi_0 \gtrsim 1$ regime, we consider photon energy close to that of the  electron beams either expected from laser-driven wakefield acceleration (for Apollon, CoReLS and {\sc zeus}) or emerging from the linear accelerators (for {\sc facet-ii} and {\sc luxe}).\\

Let us now focus on panel (a), which shows the maximum total decay probability [from Eq.~\eqref{eq:Ptot}] for a photon with energy $\gamma_\gamma$ colliding head-on with a LP Gaussian laser pulse with maximum field strength $a_0$ and duration $\tauFWHM = 25~{\rm fs}$ (typical for the facilities listed above). 
This panel suggests the strategy to optimize pair production depending on the type of facility.

First, despite the extremely high electron beam energies, facilities such as {\sc luxe} and {\sc facet-ii} (in white in Fig.~\ref{fig:fig11}) offer the possibility to probe the regime of moderate quantum parameter ($\chi_0 \lesssim 1$) for which pair production may be observed ($P_{\rm tot} \simeq 10^{-2}$) but will not be abundant. This limit might be overcome by increasing the laser power up to few 100's of TW (300 TW are envisioned at {\sc luxe}), which allows to enter the quantum regime ($\chi_0 > 1$) and increase $P_{\rm tot}$ to values exceeding $0.1$. On these facilities however, because $P_{\rm tot}$ will not assume large values, increasing $a_0$ (e.g. by focusing the laser pulse as much as technically possible) is one of the most promising path to achieve abundant pair production.

In contrast, provided that multi-GeV electron beams can be obtained from laser wakefield acceleration, abundant pair production ($P_{\rm tot} \gtrsim 0.63$) is expected on all multi-PW laser facilities with the focalisation technique already in place (typically an aperture in between $f/3$ and $f/4$ was considered in Fig.~\ref{fig:fig11}).\\

Similar conclusions can be drawn from panel (b) where we examine in more details the influence of the laser pulse spatial profile, by looking at the total cross-section\footnote{Since the Gaussian field profiles depend on the transverse coordinate through the ratio $\rho/w_{0}$ [Eq. (\ref{eq:LG_field})], $\sigma_{\rm tot}/w_{0}^{2}$ is independent of $w_{0}$.} $\sigma_{\rm tot}$ [as defined by Eq.~\eqref{eq:sigmatot}] normalized to $w_{0}^{2}$, considering a Gaussian laser beam with the same temporal properties as in panel (a).

For the {\sc facet-ii} and {\sc luxe} facilities, the total cross-section assumes very small values but increases fast as $a_0$ is increased: e.g. for {\sc luxe}, $\sigma_{\rm tot}$ increases by 3 orders of magnitude increasing $a_0$ by a factor 10. Clearly, operating at large field strength will be a bottleneck for achieving abundant pair production on these facilities.

In contrast, in the range of parameters covered by multi-PW facilities, the dependence of $\sigma_{\rm tot}$ with $a_0$ is much weaker. As in addition, $\sigma_{\rm tot} \simeq w_0^2$ (consistent with $P_{\rm tot} \simeq 1$ at the center of the beam),
it becomes interesting for these facilities to increase the laser transverse size rather than opt for tight focusing.\\

Let us finally discuss in more details the impact of focusing for the three upgrades of the Apollon facility (at 1, 3 and 7.5 PW).
In panel (c), we present the total cross section (in units of $\lambda^{-2}$) as a function of $\gamma_\gamma$, considering either the standard $f/3$ aperture (black lines) or the more challenging $f/1.5$ aperture (red lines). For each laser power, a tighter focusing reduces the beam waist but increases the maximum $a_{0}$. 

In all cases,  the cross section is rapidly increasing with the photon energy for $\gamma_{\gamma}\lesssim 2000$ ($\simeq 1$~GeV).
It is almost flat as $\gamma_{\gamma}$ reaches $10^4$ ($\simeq 5$~GeV), which suggests that it is not worth increasing the photon energy above this value in forthcoming experiments to optimize pair production in the soft shower regime.

As shown in panel (c), for a given laser power the cross section for $f/1.5$ is larger than for $f/3$ only for small values of $\gamma_{\gamma}$, while the opposite behaviour is observed at large $\gamma_{\gamma}$ (of the order of a few GeV). This means that such tight focus, which is technologically very challenging, is not necessary for pair creation in forthcoming experiments on Apollon.

To test this prediction, we have performed  complementary 3D PIC simulations of the 3 PW case with $f/3$ and $f/1.5$ ($a_0= 138.8$ and $276.7$, respectively) interacting with a gamma flash with energy $\gamma_\gamma = 10^4$. 
For the $f/3$ case we find $\sigma_{\rm tot} = 26.8\,\lambda^2$ which is about 30\% higher than the frozen case ($\sigma_{\rm tot} =22.1 ~\lambda^2$). This remains larger than the total cross-section obtained with the $f/1.5$ aperture, for which $\sigma_{\rm tot} = 19.5 ~\lambda^2$. In that case however, we note a significant (by a factor $\sim 2$) departure from the frozen case ($\sigma_{\rm tot} \simeq 9.8 ~\lambda^2$) as more secondary pairs are produced.

\section{Conclusion}\label{sec:conclusions}

In summary, we have presented a systematic study of nonlinear Breit-Wheeler pair production in the head-on collision of high-energy gamma photons with an ultra-high intensity laser beam. Combining analytical modeling and PIC simulations embarking the relevant QED modules, we  have evaluated the impact on the efficiency of pair production of the laser spatio-temporal profile [comparing e.g. Gaussian with different focal spots and Laguerre-Gaussian (LG) pulses] and parameters such as polarisation, intensity and duration. Motivated by experimental constraints, we have considered fixed laser energies while changing these parameters. 

We have explored laser field strength and photon energy parameter ranges relevant for currently and/or upcoming high-power laser facilities, focusing in particular on ultra-short Ti:Sapphire laser facilities. A reduced model was proposed that allows to describe pair production in the regime of soft-shower, where secondary pair production can be neglected. This model, that would in principle allow to predict a minimum number of produced pairs, is found to agree remarkably well with 3D PIC simulations over a broad range of parameters, highlighting the importance of the soft-shower regime for experiments on the forthcoming laser facilities. 

The model also allows to distinguish two regimes of interaction depending on whether the probability for a photon to decay into a pair as it crosses a single half-wavelength of the laser pulse is large (of order 1) or not. The two regimes are not however fully determined by the photon maximum quantum parameter $\chi_0$, so that the laser field strength parameter $a_0$ and photon energy $\gamma_\gamma$ do not play a symmetric role in the system dynamics. This was confirmed in 3D PIC simulations where the two regimes were investigated at $\chi_0=4.85$ considering laser beams with complex spatio-temporal profiles, LG beams in particular, and different polarizations. It was found that, for a fixed laser energy, using circular polarization or LG does not improve significantly pair production. As they are also quite difficult to achieve experimentally, their effect on pair production is too marginal to be interesting for applications. It was found however that, in the regime of high pair production probability, it can be preferable not to focus much the laser beam (keeping the same total energy with lower peak intensity) and maintain a Gaussian shape in order to maximise the area with high-enough fields for efficient pair production. 

These findings help to draw guidelines for future experiments on ultra-high intensity facilities. 
We show in particular that the path to abundant pair production is different whether one considers $100\,$TW-class or multi-PW laser systems.
In particular, for facilities such as {\sc facet-ii} or {\sc luxe}, 10 to $100\,$TW-class lasers will be coupled to ultra-relativistic 10-20 GeV electron beams emerging from conventional electron accelerators. On these facilities, pair production will develop in the low probability regime and operating in a tightly-focused configuration to access the highest possible laser field strength will be a major experimental challenge.
In contrast, using multiple laser beam multi-PW facilities will operate in the regime of high pair production probability. In this regime, the pair production efficiency (measured in our work in terms of a total cross section) depends more strongly on the laser pulse transverse size than on its maximal field strength so that operating with standard (not too tight) focusing aperture increases the number of produced pairs. Interestingly, producing high-energy photons (or electrons, e.g. through laser wakefield acceleration) of a few ($\sim\! 5$) GeV shall be enough for these facilities as higher energies do no help increase significantly the pair production efficiency in this high probability regime.

To conclude, this work provides a deeper understanding of the optimal conditions for pair production in upcoming experimental campaigns exploring nonlinear Breit-Wheeler pair production. It focused on pair production seeded by high-energy photons, and it will be interesting for future works to investigate the differences with the case where ultra-relativistic electrons are used to seed the pair production process. Studying the transition from the soft-shower to the cascade/avalanche regime in which secondary pairs and radiations play a dominant role is another interesting perspective.

\section*{Acknowledgements}
The authors are grateful to F. Amiranoff, P. Audebert, S. Corde, L. Gremillet, L. Lancia, R. Nuter and C. Thaury for stimulating discussions.
This work used the open-source PIC code \Smilei, the authors are grateful to all \Smilei contributors and to the \Smilei-dev team for its support. 
Simulations were performed on the Irene-Joliot-Curie machine hosted at TGCC, France, using High Performance Computing resources from GENCI-TGCC 
(Grant No. 2020-x2016057678). 
Support by Sorbonne Université in the framework of the Initiative Physique des Infinis (IDEX SUPER) is acknowledged.
This publication is also supported by the Collaborative Research Centre 1225 funded by Deutsche Forschungsgemeinschaft 
(DFG, German Research Foundation) - Project-ID 273811115 - SFB 1225.

%%%%%%%%%% APPENDIX =======================================================================================
\appendix
\section{Approximation of \texorpdfstring{$\mathcal{I}_{\varepsilon}(\chi_0)$}{TEXT} }\label{app:ComputeIntI}

In this Appendix, we detail how the integral $\mathcal{I}_{\varepsilon}(\chi_0) = \int_0^\pi b_0\!\left(\chi_0 \Psi_{\varepsilon}(\varphi)\right) d\varphi$ [Eq.~\eqref{eq:intI}] 
can be efficiently approximated by Eq.~\eqref{eq:intIApproximated}. 
To do so, let us start by stressing that, in general, and in particular at small $\chi$, $b_0(\chi)$ is a steep function of $\chi$. One thus expects the principal contribution to the integral to come from the phase $\varphi$ around $\varphi_m = \pi/2$, for which $\chi_0\,\Psi_{\varepsilon}(\varphi) \simeq \chi_m$ with $\chi_m=\chi_0/\sqrt{1+\varepsilon^2}$. 
To compute the contribution of this maximum analytically, we approximate the integrand in Eq.~\eqref{eq:intI} around $\phi_m$ as
\begin{eqnarray}\label{eq:ansatz}
b_0\!\left(\chi_0 \Psi_{\varepsilon}(\varphi)\right) \simeq b_0(\chi_m)\,\exp\left(-\frac{(\varphi-\varphi_m)^2}{2 s_{\varepsilon}^2(\chi_m)}\right)\,,
\end{eqnarray}
where $s_{\varepsilon}(\chi_m)$ is chosen so that the exact and approximated integrand have the same second derivative at $\varphi=\varphi_m$, which gives
\begin{eqnarray}
s_{\varepsilon}(\chi_m) = \sqrt{\frac{3}{2}}\,\frac{c(\chi_m)}{\sqrt{1-\varepsilon^2}} 
\quad {\rm with} \quad c(\chi)=\sqrt{\frac{2 b_0(\chi)}{3\chi b_0'(\chi)}}\,.
\end{eqnarray}
With the ansatz Eq.~\eqref{eq:ansatz}, the integral can be performed analytically, leading to
\begin{eqnarray}\label{eq:app:intIapprox}\hspace{-1cm}
\mathcal{I}_{\varepsilon}(\chi_0) \simeq  \pi\,b_0\left(\chi_m\right)\,F\!\left(s_{\varepsilon}(\chi_m)\right)
\quad {\rm with} \quad F(s)=\sqrt{2/\pi}\,s\,{\rm erf}\big(\pi\sqrt{2}/(4s)\big)\,.
\end{eqnarray}
The ansatz Eq.~\eqref{eq:ansatz} gives a good approximation of the integrand for arbitrary values of $\varepsilon$ as long as $\chi_m$ is small enough. It is also exact (for all $\chi_0$) in the cases $\varepsilon=\pm 1$ for which $s_{\pm 1}(\chi_m) \rightarrow +\infty$. Hence, Eq.~\eqref{eq:app:intIapprox} provides a very good, fully analytical approximation of Eq.~\eqref{eq:intI} for a wide range of $\varepsilon$ and $\chi_0$.

The approximation however needs to be corrected for $\chi_0 \gg 1$ and $\vert\varepsilon\vert < 1$. In this limit, the integrand in Eq.~\eqref{eq:intI} can be approximated using Eq.~\eqref{eq:b0largeChi} leading to
\begin{eqnarray}\label{eq:app:intIapproxLargeChi}\hspace{-1cm}
\mathcal{I}_{\varepsilon}(\chi_0) \simeq  \pi\,b_0\left(\chi_m\right)\,f(\varepsilon)
\quad {\rm with} \quad f(\varepsilon)=\frac{1}{\pi}\int_0^\pi \left[\sin^2\varphi+\varepsilon^2\cos^2\varphi\right]^{1/3} d\varphi\,.
\end{eqnarray}

Equations~\eqref{eq:app:intIapprox} and~\eqref{eq:app:intIapproxLargeChi} can be combined in Eq.~\eqref{eq:intIApproximated}. This form ensures the correct asymptotic behaviour of the integral Eq.~\eqref{eq:intI} in both small and large $\chi_{0}$ limits. It departs from the exact expression (integrated numerically) by less than 20$\%$, the error being maximum for intermediate values of $\chi_0 \sim 1$. This was tested over a broad range of $\chi_0 \in [0.1\, ,\, 1000]$ for various polarization parameter $\varepsilon \in \{0,0.25,0.5,0.75,1\}$.

\section{Details on 1D PIC simulations}\label{app:1Dsimulations}
A series of 1D3V (1 dimensional in space and 3 dimensional in velocity) PIC simulations, considering a plane wave colliding head-on with a flash of gamma photons, have been performed to produce Fig.~\ref{fig:Pm}c. We simulated a box of length $2.5\,\lambda$ with spatial resolution $\lambda/256$, for a simulation time of $1.75\,\tau$, with temporal resolution $\tau/512$. This is a long enough time to study the full interaction of the plane wave with a flash of gamma photons of extension $\lambda/2$ and extract the probability from the number of surviving photons. The probability map in Fig.~\ref{fig:Pm}c is obtained from the results of 4096 ($64\times64$) simulations performed over a the logarithmically spaced range of $a_{0} \in [10,4000]$ and $\gamma_{\gamma} \in [100,40000]$.

\end{document}